\documentclass{bioinfo}
\copyrightyear{2014}
\pubyear{2014}

\usepackage{url}
\usepackage{tikz}
\usetikzlibrary{arrows}
\usepackage{listings}
\usepackage{pgfplots}

\newcommand{\q}{\phantom0}
\newcommand{\qq}{\phantom{00}}

\newcommand{\qc}{\phantom{0.}}
\newcommand{\qqc}{\phantom{00.}}

\newcommand{\hdd}{$^\mathrm{HDD}$}
\newcommand{\ssd}{$^\mathrm{SSD}$}

\newcommand{\mcOT}{\multicolumn{3}{c}{\em out of time ($>10$ hours)}}
\newcommand{\mcOM}{\multicolumn{3}{c}{\em out of memory}}
\newcommand{\mcOD}{\multicolumn{3}{c}{\em out of disk ($>650$\,GB)}}
\newcommand{\mcNS}{\multicolumn{3}{c}{\em unsupported $k$}}
\newcommand{\mcF}{\multicolumn{3}{c}{\em failed}}

\begin{document}
\firstpage{1}
\title[KMC~2: Fast and resource-frugal $k$-mer counting]{KMC~2: Fast and resource-frugal $k$-mer counting}
\author[Sebastian Deorowicz, Marek Kokot, Szymon Grabowski, Agnieszka Debudaj-Grabysz]{Sebastian Deorowicz\,$^{1}\footnote{to whom correspondence should be addressed}$, Marek Kokot\,$^{1}$, Szymon Grabowski\,$^{2}$, Agnieszka Debudaj-Grabysz\,$^{1}$}
\address{$^{1}$Institute of Informatics, Silesian University of Technology, Akademicka 16, 44-100 Gliwice, Poland\\
$^{2}$Computer Engineering Department, Technical University of {\L}\'{o}d\'{z}, Al.\ Politechniki 11, 90-924 {\L}\'{o}d\'{z}, Poland}

\history{Received on XXXXX; revised on XXXXX; accepted on XXXXX}
\editor{Associate Editor: XXXXXXX}
\maketitle

\begin{abstract}
\section{Motivation:}
Building the histogram of occurrences of every $k$-symbol long substring 
of nucleotide data is a standard step in many bioinformatics applications, 
known under the name of $k$-mer counting.
Its applications include developing de Bruijn graph genome assemblers, 
fast multiple sequence alignment and repeat detection.
The tremendous amounts of NGS data require fast algorithms for 
$k$-mer counting, preferably using moderate amounts of memory.

\section{Results:}
We present a novel method for $k$-mer counting, on large datasets 
at least twice faster than the strongest competitors (Jellyfish~2, KMC~1), 
using about 12\,GB (or less) of RAM memory.
Our disk-based method bears some resemblance to MSPKmerCounter, yet 
replacing the original minimizers with signatures
(a carefully selected subset of all minimizers) and using 
$(k, x)$-mers
allows to significantly reduce the I/O, and a highly parallel 
overall architecture allows to achieve unprecedented processing speeds.
For example, KMC~2 allows to count the 28-mers of a human reads collection 
with 44-fold coverage (106\,GB of compressed size) in about 20 minutes, 
on a 6-core Intel i7 PC with an SSD.

\section{Availability:}
KMC~2 is freely available at \url{http://sun.aei.polsl.pl/kmc}.
\section{Contact:} \href{sebastian.deorowicz@polsl.pl}{sebastian.deorowicz@polsl.pl}
\end{abstract}

\section{Introduction}

One of common preliminary steps in many bioinformatics algorithms
is the procedure of {\em k-mer counting}.
This primitive consists in counting the frequencies of all $k$-long
strings in the given collection of sequencing reads,
where $k$ is usually more than 20,
and has applications in de novo assembly using de Bruijn graphs,
correcting reads and repeat detection, to name a few areas.
More applications can be found, e.g., in~\citep{MK11}, with references 
therein.\looseness=-1

$K$-mer counting is arguably one of the simplest (both conceptually
and programmatically) tasks in computational biology,
{\em if we do not care about efficiency}.
The number of existing papers on this problem suggests however 
that efficient execution of this task, with reasonable memory use,
is far from trivial.
The most successful of early approaches was Jellyfish~\citep{MK11},
maintaining a compact hash table (HT) and using lock-free operations
to allow parallel updates.
The original Jellyfish version (as presented in~\citep{MK11})
required more than 100\,GB of memory to handle human genome data
with 30-fold coverage. 
BFCounter~\citep{MP11} employs the classic
compact data structure, Bloom filter (BF), to reduce the memory
requirements due to preventing most single-occurrence $k$-mers 
(which are usually results of sequencing errors and for most applications 
can be discarded) from being added to a hash table.
Although BF is a probabilistic mechanism, BFCounter applies it in a
smart way, which does not produce counting errors.
Unfortunately, BFCounter is single-threaded and its performance
is not competitive (see also the experimental results in~\citep{DDG13}).
DSK~\citep{RLC13} and KMC~\citep{DDG13} are two disk-based algorithms.
On a high level, they are similar and partition the set of $k$-mers
into disk buckets, which are then separately processed.
DSK is more memory frugal and may process human genome data
in as little as 4\,GB of RAM, while KMC is faster but typically uses
about 11--16\,GB of RAM.
Turtle~\citep{RBS14} 
bears some similarities to BFCounter.
The standard Bloom filter is there replaced with its cache-friendly
variant~\citep{PSS09} and the hash table is replaced with a sorting
and compaction algorithm (which, accidentally, resembles a component
of KMC), apart from adding parallelism and a few smaller modifications.
Finally, MSPKmerCounter~\citep{LY14} is another disk-based algorithm,
based on the concept of minimizers, described in detail in the next
section.\looseness=-1

In this paper we present 
a new version 
of KMC, one of the fastest and most memory efficient programs.
The new release borrows from the efficient architecture of KMC~1 but reduces the disk usage several times (sometimes about 10 times) and improves the speed usually about twice.
In consequence, our tests show that KMC~2 is the fastest (by a far margin) 
algorithm for counting $k$-mers, 
with even smaller memory consumption than its predecessor.

There are two main ideas behind these improvements.
The first is the use of signatures of $k$-mers that are a generalization of the idea of \textit{minimizers}~\citep{RHY04,RHH04}.
Signatures allow significant reduction of temporary disk space.
The \textit{minimizers} were used for the first time for the $k$-mer counting in MSPKmerCounter, but our 
modification 
significantly reduces 
the main memory requirements 
(up to 3--5 times) as well as disk space (about 5~times) 
as compared to MSPKmerCounter.
The second main novelty is the use of $(k,x)$-mers ($x > 0$) for reduction 
of the amount of data to sort.
Simply speaking, instead of sorting some 
amount 
of $k$-mers we sort 
a much smaller portion of $(k+x)$-mers
and then obtain the statistics for $k$-mers in the postprocessing phase.

\begin{methods}
\section{Methods}
\label{sec:method}

\subsection{Minimizers of $k$-mers}
Most $k$-mer counting algorithms start in the same way:
they process each read from left to right and extract all $k$-mers 
from them, one by one.
Although the destination for $k$-mers 
(hash table in Jellyfish, Bloom filter in BFCounter, disk in DSK and KMC~1) 
and other details differ in particular solutions, 
the first step remains essentially the same.
There is high redundancy in such approach as consecutive $k$-mers share 
$k-1$ symbols.

An obvious idea of reducing the redundancy is to store (in some way) 
a number of consecutive $k$-mers (ideally even a complete read) in one place.
Unfortunately, to collect the statistics we need to find all copies of each unique $k$-mer, which is not an easy task when the copies are stored in many places.
A clever solution to these problems is based on the concept 
of minimizers~\citep{RHY04,RHH04}.
A \emph{minimizer} of a $k$-mer is such of its $m$-mers ($m < k$) that no other lexicographically smaller $m$-mer can be found.
The crucial observation is that 
usually many consecutive $k$-mers have the same minimizer, so in memory or 
in a file on disk they can be represented as one sequence of more than $k$ symbols, significantly reducing the redundancy.

The idea of minimizers was adopted recently for $k$-mer counting~\citep{LY14}.
Since in genomic data the read direction is rarely known, $k$-mer counters usually do not distinguish between direct $k$-mers and their reverse complements, and collect statistics for \emph{canonical} $k$-mers.
The canonical $k$-mer is lexicographically smaller of the pair: 
the $k$-mer and its reverse complement.
Therefore, Li and Yan in their MSPKmerCounter use \emph{canonical minimizers}, i.e., 
the minima of all canonical $m$-mers from the $k$-mer.
They process the reads one by one and look for 
contiguous areas containing $k$-mers having the same canonical minimizer; 
they dub these areas as ``super $k$-mers''.
Then, the resulting super $k$-mers 
are distributed into one of several \emph{bins} (disk files) according to the related canonical minimizer (more precisely, according to its hash value; 
in this way the number of resulting bins is kept within reasonable limits).
In the second stage each bin is loaded into main memory (one by one), all $k$-mers are extracted from the super $k$-mers, and then counted using 
a hash table; after processing a bin the entries from the hash 
table are dumped to disk and the hash table memory reclaimed.
Since each bin contains only a small fraction of all $k$-mers present in the input data, the amount of memory necessary to process the bin is much smaller that in the case of whole input data.

This elegant idea allows to significantly reduce the disk space compared 
to storing each $k$-mer separately (as KMC~1 and DSK do).
Unfortunately, it has the following drawbacks: 
\begin{enumerate}
\item The distribution of bin sizes is far from uniform. 
In particular, the bin associated with the minimizer AA...A is usually huge. 
Other minimizers with a few As in their prefix also tend to produce large bins.
\item When a minimizer starts with a few As, then it often implies several 
new super $k$-mers spanning a single $k$-mer only.
To given an example, with $m=7$ and AAAAAAC as the minimizer: 
when the minimizer falls off the sliding window, so the current $k$-mer 
starts with AAAAAC, then AAAAACX (for some X) will likely be the new minimizer; 
but unfortunately for yet another window AAAACXY (for some Y) also has a fair 
chance to be a minimizer, etc.
\end{enumerate}

As the amount of main memory needed by MSPKmerCounter is directly related to the number of $k$-mers in the largest bin, especially the former issue is important.
It will be shown in the experimental section that 
the file corresponding to 
the minimizer AA...A can be really large.

\subsection{From minimizers to signatures}
To overcome the aforementioned problems 
we resign from ``pure'' minimizers 
and prefer to use the term of \emph{signatures} of $k$-mers.
Essentially, a signature can be any $m$-mer of $k$-mer,
but in this paper we are interested 
in such signatures that solve both of the problems mentioned above.
Namely, good signatures of length $m$ should satisfy the following 
conditions:
\begin{enumerate}
\item The size of the largest bin should be as small as possible.
\item The number of bins should be neither too large nor too small.
\item The sum of bin sizes should be as small as possible. 
\end{enumerate}

Point 1 is obvious as it limits the maximum amount of needed memory.
Point 2 protects from costly operations on a large number of files 
(open, close, append, etc.) in case of too many bins 
but also from load balancing difficulties on a multi-core system
when the number of bins is small.
The last point refers to the disk space, so minimizing it reduces the total I/O.

Obtaining optimal signatures, i.e., such that cannot be improved in any of the
listed aspects, 
seems hard, so 
a compromise must be found.
Since the origin of both problems are runs of As (especially as signature prefixes), we propose to use as signatures canonical minimizers, but only such 
that do not start with AAA, neither start with ACA, 
neither contain AA anywhere except at their beginning.
We note that in earlier works on minimizers~\citep{RHY04,RHH04,WS14} similar  problems were spotted 
(in different applications) and somewhat different solutions were presented.


As the experiments show (cf.\ experimental section of the paper), 
such a 
modification
significantly reduces the size of the largest bin and also reduces 
the total number of super $k$-mers, 
therefore
both the main memory and temporary disk use is much smaller 
compared to 
using just canonical minimizers.

\subsection{$(k, x)$-mers}
In the memory-frugal $k$-mer counters (DSK, KMC~1, MSPKmerCounter) all the input $k$-mers are split into parts to reduce the amount of RAM memory necessary to store all the $k$-mers in explicit form.
Then, the $k$-mers are sorted, inserted into a hash table or Bloom filter.
Nevertheless, often the size of the largest part (bin) can be a problem, i.e., 
affects the peak RAM use.
Also, there is a need to explicitly process (sort, insert into some data structure) each single $k$-mer.

Below we show that it is possible to reduce the amount of memory necessary for collecting the statistics even more and also speed up the sorting process by processing a significant part of $k$-mers implicitly.
To this end, we need to introduce 
$(k, x)$-mers 
that are $(k+x')$-mers in the canonical form, where $x' = 0, 1, \ldots, x$ (for some small $x$) such that all $k$-mers within 
$(k, x)$-mer
are in canonical form.

The idea is that instead of 
breaking super $k$-mers 
into $k$-mers (for sorting purposes), 
we break them
into as few as possible $(k, x)$-mers
in such way that no two neighbors share the same $k$-mer, but each $k$-mer present 
in a super $k$-mer is present in some of $(k, x)$-mers.
As preliminary experiments on real data show, with setting $x=3$ the number 
of $(k, x)$-mers becomes about twice smaller than the number of $k$-mers.
This means that the main memory is reduced almost twice.
At the same time, the sorting speed is improved.

\subsection{Sketch of the algorithm}
Similarly to its predecessor, KMC~2 has two phases: distribution and sorting.
In the distribution phase, the reads are read from FASTQ/FASTA files.
Each read is scanned to find 
(partially overlapping) 
regions (super $k$-mers)
sharing the same signature 
(Fig.~\ref{fig:split}).
These super $k$-mers 
are sent to bins (disk files) related to signatures.
The number of possible signatures, $4^m$, can be, however, quite large, e.g., 16,384 for typical value $m=7$.
Thus, to reduce the number of bins to at most 
512,
some signatures are merged (i.e., the corresponding sequences 
are sent to the same bin).
To decide which signatures to merge, 
in a preprocessing stage KMC~2 reads a small fraction 
of the input data, builds a histogram of found signatures, 
and finally merges the least frequent signatures.

\begin{figure}
\begin{tikzpicture}[>=stealth,x=0.55cm,y=0.4cm]
\small\sffamily

\draw(7,1)[black,anchor=north] node{\bfseries\footnotesize Minimizers};
\draw(0,0)[black,anchor=north west] node{\ttfamily CGTTGATCAATTTG};
\draw(0,-1)[black,anchor=north west] node{\ttfamily{\bfseries CGTT}GATC};
\draw(0,-1.8)[black,anchor=north west] node{\ttfamily\phantom{C}GT{\bfseries TGAT}CAAT};
\draw(0,-2.6)[black,anchor=north west] node{\ttfamily\phantom{CGTT}GATC{\bfseries AATT}};
\draw(0,-3.4)[black,anchor=north west] node{\ttfamily\phantom{CGTTG}ATCA{\bfseries ATTT}G};

\draw(6.2,0)[black,anchor=north west] node{\footnotesize Read};
\draw(6.2,-1)[black,anchor=north west] node{\footnotesize Minimizer: rev\_comp({\ttfamily\small CGTT}) = {\ttfamily\small AACG}};
\draw(6.2,-1.8)[black,anchor=north west] node{\footnotesize Minimizer: rev\_comp(\ttfamily\small TGAT) = {\ttfamily\small ATCA}};
\draw(6.2,-2.6)[black,anchor=north west] node{\footnotesize Minimizer: {\ttfamily\small AATT}};
\draw(6.2,-3.4)[black,anchor=north west] node{\footnotesize Minimizer: rev\_comp({\ttfamily\small ATTT}) = {\ttfamily\small AAAT}};

\draw(7,-6)[black,anchor=north] node{\bfseries\footnotesize Signatures};
\draw(0,-7)[black,anchor=north west] node{\ttfamily CGTTGATCAATTTG};
\draw(0,-8)[black,anchor=north west] node{\ttfamily{\bfseries CGTT}GATC};
\draw(0,-8.8)[black,anchor=north west] node{\ttfamily\phantom{C}GT{\bfseries TGAT}CAAT};
\draw(0,-9.6)[black,anchor=north west] node{\ttfamily\phantom{CGTT}GATC{\bfseries AATT}TG};

\draw(6.2,-7)[black,anchor=north west] node{\footnotesize Read};
\draw(6.2,-8)[black,anchor=north west] node{\footnotesize Signature: rev\_comp({\ttfamily\small CGTT}) = {\ttfamily\small AACG}};
\draw(6.2,-8.8)[black,anchor=north west] node{\footnotesize Signature: rev\_comp({\ttfamily\small TGAT}) = {\ttfamily\small ATCA}};
\draw(6.2,-9.6)[black,anchor=north west] node{\footnotesize Signature: {\ttfamily\small AATT}};

\end{tikzpicture}
\caption{A toy example of splitting a read into super $k$-mers. 
The assumed parameters are: $k = 8$, $m = 4$.}
\label{fig:split}
\end{figure}

%

In the sorting phase, KMC~2 reads a file, extracts the 
$(k, x)$-mers
from super $k$-mers and performs radix sort algorithm on them.
Then, it calculates the statistics for $k$-mers.
In real implementation $x$ can be 0, 1, 2, or 3, 
but for presentation clarity 
we will describe how to collect the statistics of $k$-mers from 
$(k, 1)$-mers.

\begin{figure}
\begin{tikzpicture}[>=stealth,x=0.55cm,y=0.4cm]
\footnotesize
\ttfamily\small

\draw(7,1)[black,anchor=north] node{\bfseries\sffamily\footnotesize Super $k$-mer};
\draw(0,0)[black,anchor=north west] node{ACGCGACGATG{\bfseries AACT}GCCATCTCACA};

\draw(7,-1.5)[black,anchor=north] node{\bfseries\sffamily\footnotesize Successive $(k,1)$-mers};
\draw(0,-2.5)[black,anchor=north west] node{ACGCGACGATGAACT};
\draw(0,-3.3)[black,anchor=north west] node{GCAGTTCATCGTCGCG};
\draw(6.2,-3.3)[black,anchor=north west] node{{\sffamily\footnotesize rev\_comp}({\ttfamily CGCGACGATGAACTGC})};
\draw(0,-4.1)[black,anchor=north west] node{CGACGATGAACTGCCA};
\draw(0,-4.9)[black,anchor=north west] node{ACGATGAACTGCCATC};
\draw(0,-5.7)[black,anchor=north west] node{AGATGGCAGTTCATC};
\draw(6.2,-5.7)[black,anchor=north west] node{{\sffamily\footnotesize rev\_comp}(GATGAACTGCCATCT)};
\draw(0,-6.5)[black,anchor=north west] node{ATGAACTGCCATCTCA};
\draw(0,-7.3)[black,anchor=north west] node{GAACTGCCATCTCACA};

\draw(7,-8.8)[black,anchor=north] node	{\bfseries\sffamily\footnotesize Sorted $(k,1)$-mers};
\draw(0,-9.8)[black,anchor=north west] node {ACGCGACGATGAACT};
\draw(0,-10.6)[black,anchor=north west] node {AGATGGCAGTTCATC};
\draw(0,-11.4)[black,anchor=north west] node {ACGATGAACTGCCATC};
\draw(0,-12.2)[black,anchor=north west] node{ATGAACTGCCATCTCA};
\draw(0,-13.0)[black,anchor=north west] node{CGACGATGAACTGCCA};
\draw(0,-13.8)[black,anchor=north west] node{GAACTGCCATCTCACA};
\draw(0,-14.6)[black,anchor=north west] node{GCAGTTCATCGTCGCG};

\draw[decorate, decoration={brace}, line width={1pt}] (6,-10.0) -- (6,-11.4);
\draw(6.3,-10.7) [black,anchor=west] node{$R_0$};
\draw[decorate, decoration={brace}, line width={1pt}] (6,-11.6) -- (6,-13.0);
\draw(6.3,-12.3) [black,anchor=west] node{$R_A$};
\draw[decorate, decoration={brace}, line width={1pt}] (6,-13.1) -- (6,-13.9);
\draw(6.3,-13.5) [black,anchor=west] node{$R_C$};
\draw[decorate, decoration={brace}, line width={1pt}] (6,-14.0) -- (6,-15.4);
\draw(6.3,-14.7) [black,anchor=west] node{$R_G$};
\draw[decorate, decoration={brace}, line width={1pt}] (8,-11.6) -- (8,-15.4);
\draw(8.3,-13.5) [black,anchor=west] node{$R_1$};
\end{tikzpicture}
\caption{Splitting a super $k$-mer into $(k,1)$-mers followed by sorting them.
The assumed parameters are: $k = 15$, $m = 4$.
The range $R_T$ is empty (thus not shown).}
\label{fig:kx-mers}
\end{figure}

It is important to notice where in the sorted array of 
$(k, 1)$-mers
some $k$-mer can be found.
There are 6 possibilities: 
\begin{enumerate}
\item[1] it can be just a $k$-mer,
\item[2] it can be a prefix of some $(k+1)$-mer,
\item[3--6] it can be a suffix of $(k+1)$-mer preceded by A, C, G, or T.
\end{enumerate}

Therefore, we conceptually split the array of 
$(k, 1)$-mers
into 5 non-overlapping, sorted subarrays: one ($R_0$) containing $k$-mers and four ($R_A$, $R_C$, $R_G$, $R_T$) containing $(k+1)$-mers starting 
with A, C, G, T.
There is also one 
extra 
subarray ($R_1$) containing all $(k+1)$-mers, 
i.e., 
a concatenation of $R_A$, $R_C$, $R_G$ and $R_T$ (Fig.~\ref{fig:kx-mers}).

Now to collect the statistics of $k$-mers we scan these 6 subarrays in parallel.
So, we have 6 pointers somewhere in $R_*$
We compare the pointed elements, find the lexicographically smallest canonical $k$-mer among them (from $R_X$ for $X$ being a letter we take the suffix of $(k+1)$-mer) and store it in the resulting array of statistics of $k$-mers $P$ if it is different than the recently added $k$-mer to $P$. 
Otherwise, we just increase the counter related to this $k$-mer in $P$.
Since, we scan the arrays $R_*$ in a linear fashion, the time complexity of this ``merging'' subphase is linear.


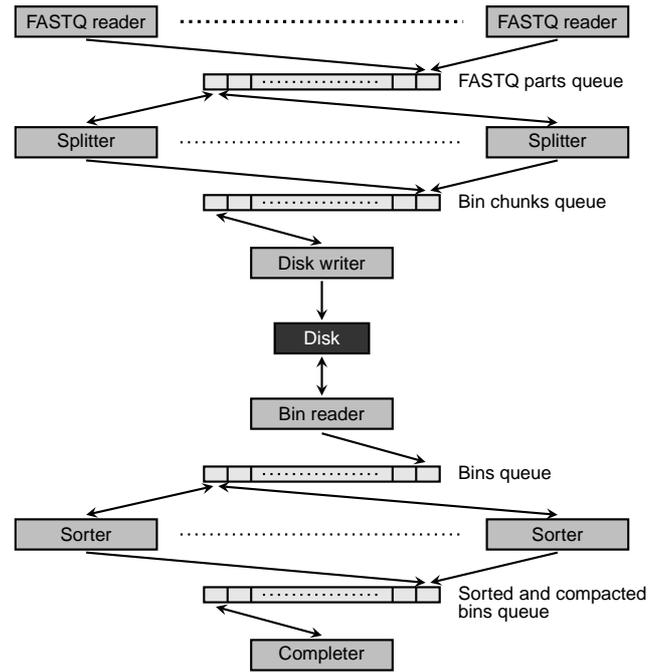
\begin{figure}
\begin{tikzpicture}[>=stealth,x=0.62cm,y=0.4cm]
\scriptsize

\filldraw[black!25!white](2,19.0) rectangle(5,20);
\draw[thick](2,19.0) rectangle(5,20);
\draw(3.5,19.5)[black] node{\sffamily FASTQ reader};
\draw[->, thick](3.5,18.90)--(10.7,17.9);

\draw[dotted,very thick](5.5,19.5)--(11.5,19.5);

\filldraw[black!25!white](12,19.0) rectangle(15,20);
\draw[thick](12,19.0) rectangle(15,20);
\draw(13.5,19.5)[black] node{\sffamily FASTQ reader};
\draw[->, thick](13.5,18.90)--(10.8,17.9);

\filldraw[black!10!white](6,17.25) rectangle(11,17.75);
\draw[thick](6,17.25) rectangle(11,17.75);
\draw[thick](6.5,17.25)--(6.5,17.75);
\draw[thick](7.0,17.25)--(7.0,17.75);
\draw[dotted, thick](7.25,17.5)--(9.75,17.5);
\draw[thick](10.0,17.25)--(10.0,17.75);
\draw[thick](10.5,17.25)--(10.5,17.75);
\draw[right](11.25,17.5) node{\sffamily FASTQ parts queue};

\filldraw[black!25!white](2,15.0) rectangle(5,16);
\draw[thick](2,15.0) rectangle(5,16);
\draw(3.5,15.5)[black] node{\sffamily Splitter};
\draw[<->, thick](3.5,16.15)--(6.2,17.10);

\draw[dotted, thick](5.5,15.5)--(11.5,15.5);

\filldraw[black!25!white](12,15.0) rectangle(15,16);
\draw[thick](12,15.0) rectangle(15,16);
\draw(13.5,15.5)[black] node{\sffamily Splitter};
\draw[<->, thick](13.5,16.15)--(6.3,17.10);

\filldraw[black!10!white](6,13.25) rectangle(11,13.75);
\draw[thick](6,13.25) rectangle(11,13.75);

\draw[thick](6.5,13.25)--(6.5,13.75);
\draw[thick](7.0,13.25)--(7.0,13.75);
\draw[dotted, thick](7.25,13.5)--(9.75,13.5);
\draw[thick](10.0,13.25)--(10.0,13.75);
\draw[thick](10.5,13.25)--(10.5,13.75);
\draw[right](11.25,13.5) node{\sffamily Bin chunks queue};

\draw[->, thick](3.5,14.90)--(10.7,13.9);
\draw[->, thick](13.5,14.90)--(10.8,13.9);

\filldraw[black!25!white](7,11.0) rectangle(10,12);
\draw[thick](7,11.0) rectangle(10,12);
\draw(8.5,11.5)[black] node{\sffamily Disk writer};
\draw[<->, thick](8.5,12.15)--(6.25,13.10);

\filldraw[black!80!white](7.5,8.5) rectangle(9.5,9.5);
\draw[thick](7.5,8.5) rectangle(9.5,9.5);
\draw(8.5,9)[white] node{\sffamily Disk};
\draw[->, thick](8.5,10.90)--(8.5,9.60);

\filldraw[black!25!white](7,6.0) rectangle(10,7);
\draw[thick](7,6.0) rectangle(10,7);
\draw(8.5,6.5)[black] node{\sffamily Bin reader};
\draw[<->, thick](8.5,7.1)--(8.5,8.4);

\filldraw[black!10!white](6,4.25) rectangle(11,4.75);
\draw[thick](6,4.25) rectangle(11,4.75);

\draw[thick](6.5,4.25)--(6.5,4.75);
\draw[thick](7.0,4.25)--(7.0,4.75);
\draw[dotted, thick](7.25,4.5)--(9.75,4.5);
\draw[thick](10.0,4.25)--(10.0,4.75);
\draw[thick](10.5,4.25)--(10.5,4.75);
\draw[right](11.25,4.5) node{\sffamily Bins queue};

\draw[->, thick](8.5,5.85)--(10.7,4.9);

\filldraw[black!25!white](2,2.0) rectangle(5,3.0);
\draw[thick](2,2.0) rectangle(5,3);
\draw(3.5,2.5)[black] node{\sffamily Sorter};
\draw[<->, thick](3.5,3.15)--(6.2,4.10);

\draw[dotted, thick](5.5,2.5)--(11.5,2.5);

\filldraw[black!25!white](12,2.0) rectangle(15,3);
\draw[thick](12,2.0) rectangle(15,3);
\draw(13.5,2.5)[black] node{\sffamily Sorter};
\draw[<->, thick](13.5,3.15)--(6.3,4.10);

\filldraw[black!10!white](6,0.25) rectangle(11,0.75);
\draw[thick](6,0.25) rectangle(11,0.75);

\draw[thick](6.5,0.25)--(6.5,0.75);
\draw[thick](7.0,0.25)--(7.0,0.75);
\draw[dotted, thick](7.25,0.5)--(9.75,0.5);
\draw[thick](10.0,0.25)--(10.0,0.75);
\draw[thick](10.5,0.25)--(10.5,0.75);
\draw[right](11.25,0.5) node{\sffamily Sorted and compacted};
\draw[right](11.25,-0.1) node{\sffamily bins queue};

\draw[->, thick](3.5,1.90)--(10.7,0.9);
\draw[->, thick](13.5,1.90)--(10.8,0.9);

\filldraw[black!25!white](7,-1.0) rectangle(10,-2);
\draw[thick](7,-1.0) rectangle(10,-2);
\draw(8.5,-1.5)[black] node{\sffamily Completer};
\draw[<->, thick](8.5,-0.85)--(6.25,0.1);

\end{tikzpicture}
\caption{A scheme of the parallel KMC algorithm}
\label{fig:scheme}
\end{figure}

The overall KMC~2 algorithm is presented in Fig.~\ref{fig:scheme}.
Several FASTQ readers send input data chunks into a queue, 
handled then by splitters which dispatch super $k$-mers with the
same signature to the same bin chunk.
The queue of these chunks is in turn processed with a disk writer, 
which dumps the bin to disk.
In the next phase, the bins, read from disk to a queue in the memory, 
are sorted and compacted by multiple sorter threads.
Finally, the completer stores the sorted bins in the output database on disk.

The final database of $k$-mers is stored in compact binary form.
The KMC~2 package contains: the $k$-mer counter, dump program that allows to produce the textual list of $k$-mers together with their counters, C++ API designed to allow to use the database directly in various applications.
The $k$-mer counter allows to specify various parameters, e.g., the threshold below which the $k$-mer is discarded (e.g., in some applications the $k$-mers appearing only once are treated as erroneous), the maximal amount of memory used in the processing.
More details on the API, the database format and the search algorithm 
in the database are given in the Supplementary material.

\subsection{Additional features}
KMC~2, like its former version, allows to refrain from counting too rare 
or too frequent $k$-mers.
It is done during ``merging'' substage, in which the total number of occurrences of each $k$-mer is known.
The software also supports quality-aware counters, compatible with 
the popular error-correction package Quake~\citep{KSS10}.
In this mode, the counter for the $k$-mer is incremented by the probability that all symbols of the $k$-mer are correct (calculated according to the base quality values).
To allow this, the qualities must be stored in temporary disk files for each base of a super $k$-mer.
To our knowledge, the only other $k$-mer counters with this functionality 
are KMC~1 and Jellyfish~1 (but not the current version 2).
KMC~2 handles not only sequencing reads (FASTQ), but also genomes (FASTA).
Finally, we note that KMC~2 can work in RAM-only mode in which the bins 
are simply stored in the main memory, which may be convenient for large 
datacenters.
\section{Results}
The implementation of KMC~2 was compared against the best, 
in terms of speed and memory efficiency, competitors: 
Jellyfish~2 (which is significantly more efficient than the version 
described in~\citep{MK11}), DSK, Turtle, MSPCounter, KAnalyze and KMC~1.
Each program was tested for two values of $k$ (28 and 55) and 
in two hardware configurations: using conventional disks (HDD) 
and using a solid-state disk (SSD).
We used several datasets (Table~\ref{tab:datasets}) of varying size; 
two of them are human data with large coverage.
The experiments were run on a machine equipped with an Intel i7 4930 CPU 
(6 cores clocked at 3.4\,GHz), 64\,GB RAM, and 2 HDDs (3\,TB each) in RAID 0 
and single SSD (1\,TB).
The programs were run with the number of threads equal to the number of 
virtual cores ($6 \times 2 = 12$), to achieve maximum speed.

\begin{table}[t]
\processtable{Characteristics of the datasets used in the experiments.\label{tab:datasets}}%
{
\renewcommand{\tabcolsep}{0.45em}
\begin{tabular}{lcccccc}\toprule
Organism						&  Genome	& No.\	& FASTQ		& No. 	& Gzipped  	& Avg. read\\
								&  length	& bases	& file size	& files	& size 		& length\\\midrule
\emph{F.\ vesca}			&	\qc210	& \qq4.5	& \q10.3		& 11		& \qq3.5		& 353	\\
\emph{G.\ gallus}			&	1,040		& \q34.7	& 115.9		& 15		& \q25.9		& 100\\
\emph{M.\ balbisiana}	&	\qc472	& \q56.9	& 197.1		& \q2		& \q49.1		& 101\\
\emph{H. sapiens} 1		&	3,093		& \q86.0	& 223.3		& \q6		& \q70.8		& 100\\
\emph{H. sapiens}	2 		& 	3,093		& 135.3	& 312.9		& 48		& 105.8		& 101\\
\botrule
\end{tabular}
}{No of bases are in Gbases. File sizes are in Gbytes (1Gbyte = $10^9$ bytes).
Approximate genome lengths are in Mbases according to \url{http://www.ncbi.nlm.nih.gov/genome/}.}
\end{table}

The comparison, presented in 
Tables~\ref{tab:res:GG}--\ref{tab:res:cox}
and Supplementary Tables 1--2, 
includes total computation time (in seconds), maximum RAM use, maximum disk use.
RAM and disk use are given in GBs (1\,GB = $2^{30}$\,B). 
Time is wall-clock time in seconds. 
A test running longer than 10 hours was interrupted.
Other reasons for not finishing a test were excessive memory consumption 
(limited by the total RAM, i.e., 64\,GB) or excessive disk use 
(over 650\,GB, chosen for our 1\,TB SSD disk; note that the largest input dataset,
\emph{H. sapiens}	2,
occupies 312.9\,GB on the same disk).

\begin{table}[!h]
\processtable{$k$-mers counting results for \emph{G. gallus}.
\label{tab:res:GG}}
{%
\renewcommand{\tabcolsep}{0.3em}
\begin{tabular}{@{\extracolsep{0.45em}}lcccp{0.4em}ccc}
\toprule
& \multicolumn{3}{c}{$k=28$} && \multicolumn{3}{c}{$k=55$}\\\cline{2-4}\cline{6-8}
Algorithm 		& RAM	& Disk& Time		&& RAM& Disk 	& Time	 \\
\midrule
\multicolumn{8}{c}{\bfseries SSD}\\
Jellyfish 2	& 33 	& \qq0& \qqc880	&& \mcOM\\
KAnalyze		& \q9	& 270	& 11,071		&& \mcNS\\
DSK			& 	\q6& 101	&\q1,325		&&	\q6&\q94	&	1,836		\\
Turtle		& 	48	&\qq0	&\q1,004		&&	\mcOM\\
MSPKC			& 17	& 114	& \q3,382	&&	\mcOT\\
KMC~1			& 13	& 101	&\qqc868		&& 12	& 173	& 	1,792		\\
KMC~2 (12GB)& 12	&\q25	& \qqc408	&& 12	&\q18	&	\qc503	\\
KMC~2	(6GB)	&\q6	&\q25	& \qqc431	&&\q6	&\q18	&	\qc562	\\
\midrule
\multicolumn{8}{c}{\bfseries HDD}\\
Jellyfish 2	& 33 	& \qq0& \qqc915	&& \mcOM\\
DSK			& \q6	& 101	&\q3,600		&&	\q6&\q94	&	4,206		\\
Turtle		& 	48	&\qq0	&\q1,058		&&	\mcOM\\
MSPKC			& 17	& 114	& \q4,853	&&	\mcOT\\
KMC~1			& 11	& 101	&\q1,320		&&	12	& 173	&	2,036		\\
KMC~2			& 12	&\q25	& \qqc587	&&	12	&\q18	&	\qc656	\\
\botrule
\end{tabular}
}{
}
\end{table}

\begin{table}[!h]
\processtable{$k$-mers counting results for \emph{M.\ balbisiana}.
\label{tab:res:mb}}
{%
\renewcommand{\tabcolsep}{0.3em}
\begin{tabular}{@{\extracolsep{0.45em}}lcccp{0.4em}ccc}
\toprule
& \multicolumn{3}{c}{$k=28$} && \multicolumn{3}{c}{$k=55$}\\\cline{2-4}\cline{6-8}
Algorithm 	& RAM	& Disk& Time		&& RAM& Disk 	& Time	 \\
\midrule
\multicolumn{8}{c}{\bfseries SSD}\\
Jellyfish 2	& 17	& \qq0& \q1,080	&& 26	&\qq0	&	\qc853	\\
KAnalyze		& \q9	& 354	&\q8,249		&&---	&---	&	---		\\
DSK			& \q6	& 164	&\q2,356		&&	\q6& 138	&	2,962		\\
Turtle		& 46	& \qq0&\q1,484		&& \mcOM\\
MSPKC			& 10	& 185	&\q8,729		&&	\mcOT\\
KMC~1			& 13	& 165	&\q1,229		&&	15	& 279	& 2,622		\\
KMC~2 (12GB)& 12	&\q41	& \qqc755	&& 12	&\q29	&	\qc834	\\
KMC~2	(6GB)	& \q6	&\q41	& \qqc685	&&\q6	&\q29	&	\qc895	\\
\midrule
\multicolumn{8}{c}{\bfseries HDD}\\
Jellyfish 2	& 17	& \qq0&\q1,115		&& 26	&\qq0	&\qc881		\\
DSK			& \q6	& 164	&\q6,216		&&\q6	& 138	&	7,228		\\
Turtle		& 46	& \qq0&\q1,498		&& \mcOM\\
MSPKC			& 10	& 185	&12,152		&&	\mcOT\\
KMC~1			& 13	& 165	&\q2,194		&&	15	& 279	& 3,367		\\
KMC~2			& 12	&\q41	& \qqc960	&&	12	&\q29	& 1,041		\\
\botrule
\end{tabular}
}{
}
\end{table}

\begin{table}[!h]
\processtable{$k$-mers counting results for \emph{H.\ sapiens} 2.
\label{tab:res:cox}}
{%
\renewcommand{\tabcolsep}{0.3em}
\begin{tabular}{@{\extracolsep{0.45em}}lcccp{0.4em}ccc}
\toprule
& \multicolumn{3}{c}{$k=28$} && \multicolumn{3}{c}{$k=55$}\\\cline{2-4}\cline{6-8}
Algorithm 	& RAM	& Disk& Time		&& RAM& Disk 	& Time	 \\
\midrule
\multicolumn{8}{c}{\bfseries SSD}\\
Jellyfish 2	& 62	& \qq0& \q3,212	&& \mcOM\\
KAnalyze		& \mcOD		 				&&	\mcNS\\
DSK			& \q6	& 263	&	\q5,487	&&\q6	&256	&	\q7,732	\\
Turtle		& \mcOM						&& \mcOM						\\
MSPKC			& \mcOT						&&	\mcOT\\
KMC~1			& 17	&396	&\q2,998		&&	\mcOD						\\
KMC~2	(12GB)& 12	&101	&\q1,615		&&	13	&\q70	&\q2,038		\\
KMC~2 (6GB)	&\q6	&101	&\q1,706		&&	13	&\q70	&\q2,446		\\
\midrule
\multicolumn{8}{c}{\bfseries HDD}\\
Jellyfish 2	& 62	& \qq0& \q3,231		&& \mcOM\\
DSK			& \q6	&	263	&	18,493	&&	\q6& 256		&	22,432\\
KMC~1			& 17	&396	&\q4,898		&&	\mcOD	\\
KMC~2			& 12	&101	&\q2,259		&&	13	&\q70	&\q2,640		\\
\botrule
\end{tabular}
}{
}
\end{table}

Several conclusions can easily be drawn from the presented tables.
Two of the competitors, KAnalyze and MSPKC, are clearly the slowest; 
for this reason, KAnalyze was tested only on the SSD.
KAnalyze also uses a large amount of temporary disk space, 
which was the reason we stopped its execution on the two human datasets 
(for $k = 28$ only, as KAnalyze does not support large values of $k$).
MSPKC, on the other hand, theoretically allows the parameter $k$ to exceed 32, 
but in none of our datasets it finished its work for $k = 55$; 
for the smallest dataset (\emph{F.\ vesca}) it failed probably because 
of variable-length reads, on the other datasets we stopped it after 
more than 10 hours of processing.
The only asset 
of KAnalyze and MSPKC we have found is their moderate 
memory use.

DSK is not very fast either.
Still, it consistently uses the smallest amount of memory 
(6\,GB was always reported) and is quite robust, as it passed all the tests.

Jellyfish~2 is not very frugal in memory use, and this is the reason 
on our machine it passed the test for $k = 55$ only for 
two datasets (\emph{F.\ vesca} and \emph{M.\ balbisiana}).
Still, for $k = 28$ it passed all the tests, being one of the fastest 
programs, often outperforming KMC~1.

Turtle is rather fast as well (slower than Jellyfish though), but even more 
memory hungry; we could not have run it on the two largest datasets.
Turtle and Jellyfish are memory-only algorithms,
all the other ones are disk-based. 
This is the reason why changing HDD to a much faster SSD does not affect 
the performance of these two counters significantly (yet it is non-zero 
due to faster input reading from the SSD).

KMC~2 on the SSD was tested twice for each $k$: 
with standard memory use (12\,GB) and with reduced memory use (6\,GB).
These settings are a ``suggestion'' rather than a rigid limitation, 
as a large maximum bin size may force KMC~2 to use more memory, 
Such a phenomenon was seen several times especially in the memory-reduced 
runs.
This also means that our goal to match DSK in memory use in 
the memory-reduced mode was not quite accomplished, 
yet 
we note that reducing the memory 
resulted in processing time longer by only 5\%--20\%.

KMC~2 with its standard memory use is a clear winner in processing time, 
on the human datasets being 
about 
twice faster than Jellyfish~2 or KMC~1.
These speed differences concern the SSD experiments, as on the HDD the gap 
diminishes (but is still significant).
This can be explained by I/O (especially reading the input data) 
being the bottleneck in several phases of KMC~2 processing.

It is worth examining how switching a conventional disk to a SSD affects 
the performance of disk-based software.
It might seem natural that the biggest time reduction (in absolute time, 
not percentage gain) should be seen in those programs which use more disk space.
To some degree it is true (e.g., KMC~1 gains more than KMC~2) 
but DSK is a ``counter-example'': e.g., on 
\emph{H. sapiens} 2 
it gains a whopping 
13,006\,s which is almost seven times the reduction for KMC~1, 
seemingly surprising as DSK uses less disk space.
Yet, a probable explanation is that DSK works in several passes, 
so its total I/O is actually quite large for large datasets.

Interestingly, for disk-based algorithms the disk use of KMC~2 is 
typically reduced when switching from $k = 28$ to $k = 55$.
This can be explained by a smaller number of $k$-mers per read, 
and in case of KMC~2 also by a smaller number of super $k$-mers per read.

We also measured how the input format (raw, gzipped) and media (HDD, SSD) 
affects the performance of our solution 
on the largest dataset, 
\emph{H. sapiens} 2 (Table~\ref{tab:res:gzipped}).
As expected, using the SSD reduces the time by 
25\%--40\%, 
and reading the input from compressed form also has a visible positive impact.
We note in passing that replacing gzip with, e.g., bzip2 (results not shown here) 
would not be a wise choice, since the improvement in compression cannot offset 
much slower bzip2's decompression.



\begin{table}[!h]
\processtable{Influence of input data format on the $k$-mers counting times of KMC~2 for \emph{H.\ sapiens} 2.
\label{tab:res:gzipped}}
{%
\renewcommand{\tabcolsep}{0.3em}
\begin{tabular}{@{\extracolsep{0.45em}}lcccp{0.4em}ccc}
\toprule
& \multicolumn{3}{c}{$k=28$} && \multicolumn{3}{c}{$k=55$}\\\cline{2-4}\cline{6-8}
Algorithm 	& RAM	& Disk& Time		&& RAM& Disk 	& Time	 \\
\midrule
\multicolumn{8}{c}{\bfseries Non-gzipped input files}\\
KMC~2\hdd	& 12	& 101	&	2,259		&&	13	& 70	&	2,640	\\
KMC~2\ssd	& 12	& 101	& 	1,615		&& 13	& 70	&	2,038	\\
KMC~2\ssd	&\q6	& 101	& 	1,706		&& 13	& 70	&	2,446	\\
\midrule
\multicolumn{8}{c}{\bfseries Gzipped input files}\\
KMC~2\hdd	& 12	& 101	&	2,004		&&13	& 70	&	2,495		\\
KMC~2\ssd	& 12	& 101	& 	1,217		&&13 	& 70	&	1,607		\\
KMC~2\ssd	&\q7	& 101	&	1,495 	&&13	& 70	&	1,909		\\
\botrule
\end{tabular}
}{}
\end{table}

Table~\ref{tab:sig_vs_min} compares signatures with minimizers on \emph{G.\ gallus}. 
We can see that using our signatures diminishes the average number of super $k$-mers in a read 
by about 10--15 percent.
Also the number of $k$-mers in the largest (disk) bin is significantly reduced, sometimes more than twice.
These achievements directly translate to smaller RAM and disk space consumption.


\begin{table}[!h]
\processtable{Comparison of signatures and minimizers for \emph{G.\ gallus} dataset.
\label{tab:sig_vs_min}}
{%
\renewcommand{\tabcolsep}{0.2em}
\begin{tabular}{@{\extracolsep{0.35em}}cp{0.0em}cccp{0.0em}ccc}
\toprule
&& \multicolumn{3}{c}{Minimizers} && \multicolumn{3}{c}{Signatures}\\\cline{3-5}\cline{7-9}
Length 	&& Avg.\ in	& No.\ $k$-mers	& Min.  	&& Avg.\ in	& No.\ $k$-mers 	& Min.\\
		 	&&  read		& largest bin 		& memory	&& 	read 	& largest bin		& memory\\
\midrule
\multicolumn{9}{c}{$k=28$}\\
5	&& 6.935	& 3,361	& 26.5	&&	6.045	& 1,904	& 18.1	\\
6	&& 7.519	& 1,231	& 10.9	&& 6.385	&\qc625	&\q5.9 	\\
7	&& 7.919	&\qc641	&\q5.5 	&& 6.728	&\qc283	&\q2.6	\\
8	&& 8.304	&\qc371	&\q3.1	&& 7.143	&\qc328	&\q3.0	\\
\midrule
\multicolumn{9}{c}{$k=55$}\\
5	&& 2.669	& 3,940	& 62.0	&& 2.477	& 2,257	& 38.3	\\
6	&& 2.915	& 1,513	& 24.7	&& 2.591	&\qc819	& 13.9	\\
7	&& 3.038	&\qc801	& 12.8	&& 2.642	&\qc280	&\q5.5	\\
8	&& 3.117	&\qc467	&\q7.3	&& 2.678	&\qc330	&\q6.4	\\
\botrule
\end{tabular}
}{`Avg.\ in read' is the average no.\ of super $k$-mers per read.
`No.\ $k$-mers largest bin' is the number (in millions) of $k$-mers in the largest bin.
`Min.\ memory' is the amount of memory (in Gbytes) necessary to process the $k$-mers in the largest bin, i.e., the lower bound of the memory requirements.
The size of temporary disk space is determined by the average number of minimizers/signatures in a read.
For example, the disk space requirements for minimizer/signature length~7 are: 25.4\,GB (signatures, $k=28$), 28.6\,GB (minimizers, $k=28$).
}
\end{table}



How $(k, x)$-mers affect bin processing is shown in Table~\ref{tab:res:kx-mers} 
for two datasets.
It is easy to see that the number of strings to sort is more than halved 
for $x = 3$, yet the speedup is more moderate, due to the extra split phase 
and sorting over longer strings.
Still, $(k, 3)$-mers vs. plain $k$-mers reduce the total time by more 
than 20\% (and even 38\% for {\em H.\ sapiens} 2 and $k = 55$).

\begin{table}[!h]
\processtable{Impact of $(k,x)$-mers on bin processing and overall KMC~2 processing, for \emph{G.\ gallus} and \emph{H.~sapiens} 2. 12\,GB RAM set, gzipped input.
``Sorted fraction'' is the ratio of the number of $(k, x)$-mers to the number 
of $k$-mers.
\label{tab:res:kx-mers}}
{%
\renewcommand{\tabcolsep}{0.3em}
\begin{tabular}{@{\extracolsep{0.25em}}cp{0.2em}ccccp{0.2em}cccc}
\toprule
&& \multicolumn{4}{c}{$k=28$} && \multicolumn{4}{c}{$k=55$}\\\cline{3-6}\cline{8-11}
$x$&& Split & Sort & Total	& Sorted		&& Split & Sort 	& Total	& Sorted		\\
	&& time 	& time & time	& fraction	&& time	& time 	& time	& fraction 	\\
\midrule
\multicolumn{11}{c}{\bfseries \emph{G.\ gallus}}\\
0 	&& 102	& 159	& 261		& 1.000		&& \q98	& 381		& 479		& 1.000		\\
1 	&& 127	& 131	& 258		& 0.646		&& 104	& 284		& 388		& 0.639		\\
2 	&& 127	& 119	& 246		& 0.539		&& 104	& 265		& 369		& 0.527		\\
3 	&& 127	& 112	& 239		& 0.491		&& 106	& 240		& 346		& 0.479		\\
\midrule
\multicolumn{11}{c}{\bfseries \emph{H.\ sapiens} ERA015743}\\
0	&& 672	& 867	& 1539	& 1.000		&& 399	& 2188	& 2587	& 1.000		\\
1	&& 664	& 669	& 1333	& 0.648		&& 448	& 1480	& 1928	& 0.638		\\
2	&& 644	& 614	& 1258	& 0.541		&& 455	& 1176	& 1630	& 0.526		\\
3	&& 644	& 573	& 1217	& 0.495		&& 439	& 1168	& 1607	& 0.478		\\
\botrule
\end{tabular}
}{For \emph{H.\ sapiens} 2 the largest bin was too large to fit the assumed amount of RAM in two cases, and the RAM consumption of KMC 2 was 25\,GB for $(55,0)$-mers, 18\,GB for $(55,1)$-mers, 15\,GB for $(55,2)$-mers, and 13\,GB for $(55,3)$-mers.}
\end{table}

The impact of $k$ on processing time and disk space is presented in Figures~\ref{fig:res:time_k} 
and~\ref{fig:res:disk_k}, respectively.
Longer $k$-mers result in even longer super $k$-mers, which minimizes I/O, but makes the sorting phase longer.
For this reason, the disk space consumption shrinks smoothly with growing $k$ (Fig.~\ref{fig:res:disk_k}), 
but the effect on processing time (Fig.~\ref{fig:res:time_k}) is not so clear. 
Still, counting $k$-mers for $k \geq 32$ is generally slower than for smaller values of $k$.

From Fig.~\ref{fig:res:disk_ram} we can see that using more memory 
accelerates KMC~2, but the effect is mediocre 
(only about 10\% speedup when raising the memory consumption from 16\,GB to 40\,GB).
The reasons behind the speedup are basically 2-fold:
$(i)$ the extra RAM allows to use 
a larger number of sorter threads (which is more efficient than 
few sorters with more internal threads per sorter), 
and $(ii)$ occasional large bins disallow to run other sorters 
at the same time if memory is limited.

Finally, we analyze the scalability and CPU load of our software (Fig.~\ref{fig:res:cpu_ram}). 
As expected, the highest speed is achieved when the number of threads
matches the number of (virtual) CPU cores (12).
Still, the time reduction between 1 and 12 threads is only by factor 3 or less,
when the input data are in non-compressed FASTQ.
Using the compressed input broadens the gap to factor 6.4 for $k = 28$ and
4.9 for $k = 55$. 
The corresponding gaps between 1 and 6 threads (i.e., equal to the number 
of {\em physical} cores) are: 
2.3 and 2.5 ($k = 28$ and $k = 55$) with non-compressed input, 
and 4.9 and 3.9 ($k = 28$ and $k = 55$) with gzipped input.
The latter experiment tells more about the scalability of our tool, 
since the performance boost from Intel hyper-threading technology can be 
hard to predict, varying from less than 10\%~\citep[Tab.~1]{SPBBXC13} 
to about 60\%~\citep[Tab.~II]{SER12} in real code.

%

\begin{figure}[t]
\centering
\pgfplotsset{width=8.0cm, height=5.25cm}
{\sffamily
\begin{tikzpicture}
\begin{axis}[
	xlabel={$k$}, 
	ylabel={Time [s]},
	minor y tick num=4,
	minor x tick num=9,
	xmin=20,
	ymin=000,
	ymax=2500,
	xmax=72,
	grid=major,
	mark size=1.5,
	font=\scriptsize,
	legend entries={12\,GB RAM, 24\,GB RAM},
	legend columns=1,
	legend pos={south east},
	legend style={font=\scriptsize, mark size=1.5},
]
\addplot[black, mark=*] table[x=k, y=m12] {time_k.dat};
\addplot[black, mark=square] table[x=k, y=m24] {time_k.dat};
\end{axis}
\end{tikzpicture}
}
\caption{Dependence of KMC~2 processing time on $k$ for \emph{H.\ sapiens} 2 dataset ($k=22, 25, 28, 32, 40, 50, 60, 70$)}
\label{fig:res:time_k}
\end{figure}
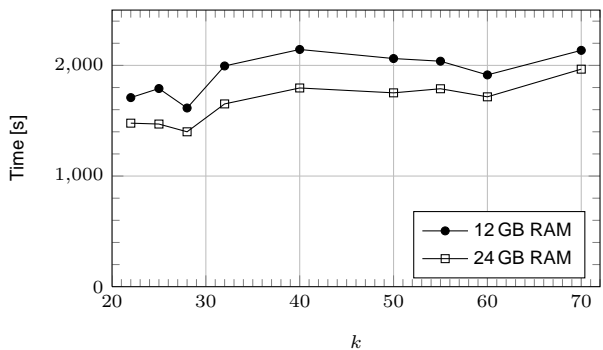

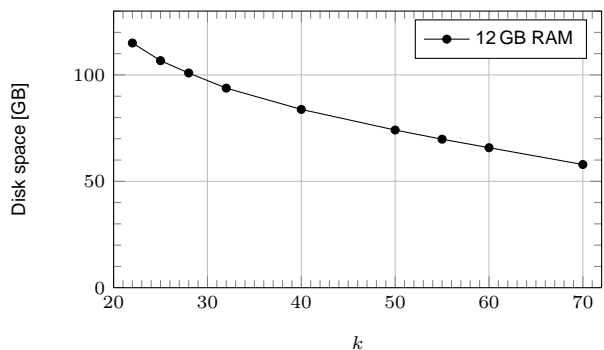
\begin{figure}
\centering
\pgfplotsset{width=8.0cm, height=5.25cm}
{\sffamily
\begin{tikzpicture}
\begin{axis}[
	xlabel={$k$}, 
	ylabel={Disk space [GB]},
	minor y tick num=4,
	minor x tick num=9,
	xmin=20,
	ymin=0,
	ymax=130,
	xmax=72,
	grid=major,
	mark size=1.5,
	font=\scriptsize,
	legend entries={12\,GB RAM},
	legend columns=-1,
	legend pos={north east},
	legend style={font=\scriptsize, mark size=1.5},
]
\addplot[black, mark=*] table[x=k, y=m12_space] {time_k.dat};
\end{axis}
\end{tikzpicture}
}
\caption{Dependence of KMC~2 temporary disk usage on $k$ for \emph{H.\ sapiens} 2 dataset}
\label{fig:res:disk_k}
\end{figure}

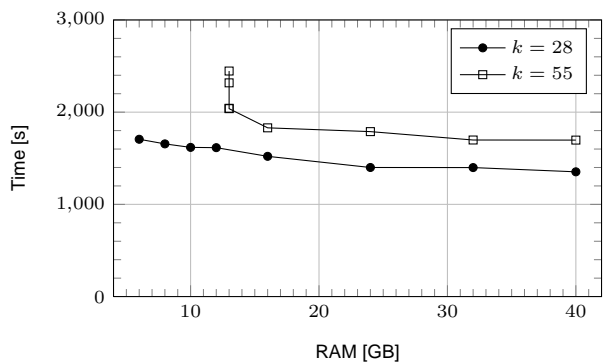
\begin{figure}
\centering
\pgfplotsset{width=8.0cm, height=5.25cm}
{\sffamily
\begin{tikzpicture}
\begin{axis}[
	xlabel={RAM [GB]}, 
	ylabel={Time [s]},
	minor y tick num=4,
	minor x tick num=9,
	xmin=4,
	ymin=000,
	ymax=3000,
	xmax=42,
	grid=major,
	mark size=1.5,
	font=\scriptsize,
	legend entries={$k=28$, $k=55$},
	legend columns=1,
	legend pos={north east},
	legend style={font=\scriptsize, mark size=1.5},
]
\addplot[black, mark=*] table[x=ram28, y=t28] {time_ram.dat};
\addplot[black, mark=square] table[x=ram55, y=t55] {time_ram.dat};
\end{axis}
\end{tikzpicture}
}
\caption{Dependence of KMC~2 processing time on maximal available RAM and type of disk for \emph{H.\ sapiens} 2 dataset.
There are 4 results for $k=55$ and 13\,GB RAM.
These results are for set 6\,GB, 8\,GB, 10\,GB, 12\,GB as maximal RAM usage.
However, the largest bin enforced to spend at least 13\,GB of RAM}
\label{fig:res:disk_ram}
\end{figure}

\begin{figure}
\centering
\pgfplotsset{width=8.0cm, height=5.5cm}
{\sffamily
\begin{tikzpicture}
\begin{axis}[
	xlabel={No.\ threads}, 
	ylabel={Time [s]},
	minor y tick num=4,
	minor x tick num=1,
	xmin=0,
	ymin=0,
	ymax=2000,
	xmax=12.5,
	grid=major,
	mark size=1.5,
	font=\scriptsize,
	legend entries={$k=28$, $k=55$, $k=28$-gz, $k=55$-gz},
	legend columns=2,
	legend pos={north east},
	legend style={font=\scriptsize, mark size=1.5},
]
\addplot[black, mark=triangle*] table[x=thr, y=t28] {time_thr.dat};
\addplot[black, mark=square*] table[x=thr, y=t55] {time_thr.dat};
\addplot[black, mark=triangle] table[x=thr, y=t28gz] {time_thr.dat};
\addplot[black, mark=square] table[x=thr, y=t55gz] {time_thr.dat};
\end{axis}
\end{tikzpicture}
\\
\bigskip
\begin{tikzpicture}
\begin{axis}[
	xlabel={No.\ threads}, 
	ylabel={CPU usage [\%]},
	minor y tick num=4,
	minor x tick num=1,
	xmin=0,
	ymin=0,
	ymax=1070,
	xmax=12.5,
	grid=major,
	mark size=1.5,
	font=\scriptsize,
	legend entries={$k=28$, $k=55$, $k=28$-gz, $k=55$-gz},
	legend columns=1,
	legend pos={north west},
	legend style={font=\scriptsize, mark size=1.5},
]
\addplot[black, mark=triangle*] table[x=thr, y=cpu28] {time_thr.dat};
\addplot[black, mark=square*] table[x=thr, y=cpu55] {time_thr.dat};
\addplot[black, mark=triangle] table[x=thr, y=cpu28gz] {time_thr.dat};
\addplot[black, mark=square] table[x=thr, y=cpu55gz] {time_thr.dat};
\end{axis}
\end{tikzpicture}
}
\caption{Dependence of KMC~2 processing time and CPU usage on the set number of threads for \emph{G.\ gallus} dataset.}
\label{fig:res:cpu_ram}
\end{figure}
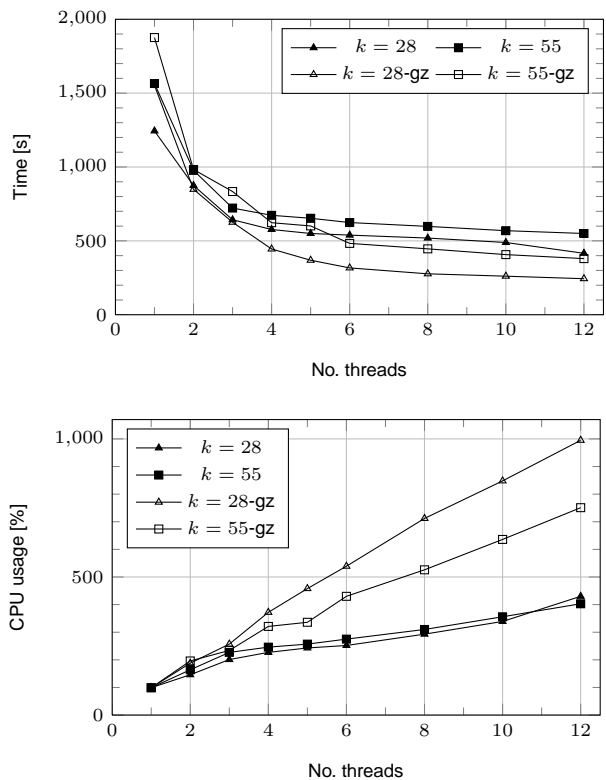

\end{methods}

\section{Conclusion}

Although the dominating trend in IT solutions nowadays is the cloud, 
the progress in bioinformatic algorithms shows that even home computers, 
equipped with multi-core CPUs, several gigabytes of RAM and a few fast 
hard disks (or one SSD disk) get powerful enough to be applied 
for real ``omics'' tasks, if their resources are loaded appropriately.

The presented KMC~2 algorithm is currently the fastest $k$-mer counter, 
with modest resource (memory and disk) requirements.
Although the used approach is similar to the one from MSPKmerCounter, 
we obtain an order of magnitude faster processing, due to the following 
KMC features: replacing the original minimizers with signatures (a carefully 
selected subset of all minimizers), using 
$(k, x)$-mers
and a highly parallel overall architecture.
As opposed to most competitors, 
KMC~2 worked stably across a large range of datasets and test settings.

In real numbers, we show that it is possible to count the 28-mers 
of a human reads collection with 44-fold coverage (106\,GB of compressed size) 
in about 20 minutes, on a 6-core Intel Core i7 PC with an SSD.
With enough amounts of available RAM it is also possible to run KMC~2 
in memory only.
In our preliminary tests it almost did not help compared to an SSD 
(up to 5\% speedup) but may be an option in datacenters, with plenty of RAM 
but possibly using network HDDs with relatively low transfer.
In this scenario a memory-only mode should be attractive.

We expect to successfully apply KMC~2 for a few 
problems related to $k$-mers, 
e.g., finding nullomers~\citep{FFBTL14}.

\section*{Acknowledgment}

\paragraph{Funding\textcolon} 
This work was supported by the Polish National Science Centre under the project DEC-2012/05/B/ST6/03148.
The work was performed using the
infrastructure supported by POIG.02.03.01-24-099/13 grant:
``GeCONiI--Upper Silesian Center for Computational Science and
Engineering''.


\begin{thebibliography}{}
\scriptsize

\bibitem[Deorowicz {\em et~al.}(2013)Deorowicz, Debudaj-Grabysz, and Grabowski]{DDG13}
Deorowicz, S., Debudaj-Grabysz, A., and Grabowski, S. (2013).
\newblock Disk-based $k$-mer counting on a PC.
\newblock {\em BMC Bioinformatics\/}, {\bf 14}, 160.

\bibitem[Falda {\em et~al.}(2014)Falda, Fontana, Barzon, Toppo, and Lavezzo]{FFBTL14}
Falda, M., Fontana, P., Barzon, L., Toppo, S., and Lavezzo, E. (2014).
\newblock keeSeek: searching distant non-existing words in genomes for {PCR}-based applications.
\newblock {\em Bioinformatics\/}, doi:10.1093/bioinformatics/btu312.

\bibitem[Kelley {\em et~al.}(2010)Kelley, Schatz, and Salzberg]{KSS10}
Kelley, D.~R., Schatz, M.~C., and Salzberg S.~L. (2010).
\newblock Quake: quality-aware detection and correction of sequencing errors.
\newblock {\em Genome Biology\/}, {\bf 11}(11):R116.

\bibitem[Kurtz {\em et~al.}(2008)Kurtz, Narechania, Stein, and Ware]{KNSW08}
Kurtz, S., Narechania, A., Stein, J., and Ware, D. (2008).
\newblock A new method to compute {K}-mer frequencies and its application to annotate large repetitive plant genomes.
\newblock {\em BMC Genomics\/}, {\bf 9}(1), 517.

\bibitem[Kurtz {\em et~al.}(2004)Kurtz, Phillippy, Delcher, Smoot, Shumway, Antonescu, and Salzberg]{KPDSSAS08}
Kurtz, S., Phillippy, A., Delcher, A.~L., Smoot, M., Shumway, M., Antonescu, C., and Salzberg, S.~L. (2004)
\newblock Versatile and open software for comparing large genomes. 
\newblock {\em Genome Biology\/}, {\bf 5}(2), R12.

\bibitem[Li and Yan(2014)Li and Yan]{LY14}
Li, Y. and Yan, X. (2014).
\newblock MSPKmerCounter: A fast and memory efficient approach for $k$-mer counting. 
Preprint at \url{http://cs.ucsb.edu/~yangli/papers/MSPKmerCounter.pdf}.

\bibitem[Mar\c{c}ais and Kingsford(2011)Mar\c{c}ais and Kingsford]{MK11}
Mar\c{c}ais, G. and Kingsford, C. (2011).
\newblock A fast, lock-free approach for efficient parallel counting of occurrences of k-mers. 
\newblock {\em Bioinformatics\/}, {\bf 27}(6), 764--770.

\bibitem[Melsted and Pritchard(2011)Melsted and Pritchard]{MP11}
Melsted, P. and Pritchard, J.~K. (2011).
\newblock Efficient counting of $k$-mers in {DNA} sequences using a {B}loom {F}ilter. 
\newblock {\em BMC Bioinformatics\/}, {\bf 12}(333).

\bibitem[Miller {\em et~al.}(2008)Miller, Delcher, Koren, Venter, Walenz, Brownley, Johnson, Li, Mobarry, and Sutton]{MDKVWBJLMS08}
Miller, J.~R., Delcher, A.~L., Koren, S., Venter, E., Walenz, B., Brownley, A., Johnson, J., Li, K., Mobarry, C.~M., and Sutton, G.~G. (2008).
\newblock Aggressive assembly of pyrosequencing reads with mates. 
\newblock {\em Bioinformatics\/}, {\bf 24}(24), 2818--2824.

\bibitem[Putze {\em et~al.}(2009)Putze, Sanders, and Singler]{PSS09}
Putze, F., Sanders, P., and Singler, J. (2009).
\newblock Cache-, hash-and space-efficient {B}loom filters.
\newblock {\em ACM Journal of Experimental Algorithms\/}, {\bf 14}.

\bibitem[Rizk {\em et~al.}(2013)Rizk, Lavenier, and Chikhi]{RLC13}
Rizk, G., Lavenier, D., and Chikhi, R. (2013).
\newblock DSK: $k$-mer counting with very low memory usage. 
\newblock {\em Bioinformatics\/}, {\bf 29}(5), 652--653.

\bibitem[Roberts {\em et~al.}(2004b)Roberts, Hayes, Hunt, Mount, and Yorke]{RHH04}
Roberts, M., Hayes, W., Hunt, B.~R., Mount, S.~M., and  Yorke, J.~A. (2004).
\newblock Reducing storage requirements for biological sequence comparison. 
\newblock {\em Bioinformatics\/}, {\bf 20}(18), 3363--3369.

\bibitem[Roberts {\em et~al.}(2004a)Robert, Hunt, Yorke, Bolanos, and Delcher]{RHY04}
Roberts, M., Hunt, B.~R., Yorke, J.~A., Bolanos, R.~A., and Delcher, A.~L. (2004).
\newblock A preprocessor for shotgun assembly of large genomes.
\newblock {\em Journal of Computational Biology\/}, {\bf 11}(4), 734--752.

\bibitem[Roy {\em et~al.}(2014)Roy, Bhattacharya, and Schliep]{RBS14}
Roy, R.~S., Bhattacharya, D., and Schliep, A. (2014).
\newblock Turtle: Identifying frequent $k$-mers with cache-efficient algorithms. 
\newblock {\em Bioinformatics\/}, doi:10.1093/bioinformatics/btu132.

\bibitem[Satish {\em et~al.}(2010)Satish, Kim, Chhugani, Nguyen, Lee, Kim, and Dubey]{SKCNLKD10}
Satish, N., Kim, C., Chhugani, J., Nguyen, A.D., Lee, V.W., Kim, D., and Dubey, P. (2010).
\newblock Fast Sort on CPUs and GPUs. A Case for Bandwidth Oblivious SIMD Sort. 
\newblock {\em Proc.\ of the 2010 Int.\ Conf.\ on Management of data\/}, pp.\ 351--362.

\bibitem[Schuepbach {\em et~al.}(2013)Schuepbach, Pagni, Bridge, Bougueleret, Xenarios, and Cerutti]{SPBBXC13}
Schuepbach, T., Pagni, P., Bridge, A., Bougueleret, L., Xenarios, I., and Cerutti, L. (2013). 
\newblock pfsearchV3: a code acceleration and heuristic to search PROSITE profiles.
\newblock {\em Bioinformatics\/}, {\bf 29}(9), 1215--1217. 

\bibitem[Sebasti{\~a}o {\em et~al.}(2012)Sebasti{\~a}o, Encarna{\c{c}}{\~a}o, and Roma]{SER12}
Sebasti{\~a}o, N., Encarna{\c{c}}{\~a}o, G., and Roma, N. (2012).
\newblock Implementation and performance analysis of efficient index structures for DNA search algorithms in parallel platforms.
\newblock {\em Concurrency and Computation: Practice and Experience\/}, doi: 10.1002/cpe.2970.

\bibitem[Wood and Salzberg(2014)Wood and Salzberg]{WS14}
Wood, D.~E., Salzberg, S.~L. (2014). 
\newblock Kraken: Ultrafast metagenomic sequence classification using exact alignments. 
\newblock {\em Genome Biology\/}, {\bf 15}(3), R46.

\end{thebibliography}

\newpage
\onecolumn
\setcounter{section}{0}
\centerline{\bfseries\Large Supplementary material for the paper}
\bigskip
\centerline{\bfseries\LARGE\itshape KMC~2: Fast and resource-frugal $k$-mer counting}
\bigskip
\centerline{\bfseries\Large by}
\bigskip
\centerline{\bfseries\Large Sebastian Deorowicz, Marek Kokot, Szymon Grabowski,}
\medskip
\centerline{\bfseries\Large and Agnieszka Debudaj-Grabysz}
\bigskip
\bigskip
\bigskip
\bigskip


\clearpage
\section{KMC usage}
\label{sec:usage}

\textsf{KMC} 2 program constructs a database of statistics for input set of FASTQ files.
This database can then be used from other software: directly via KMC API (described in Section~\ref{sec:api}) or by reading a textual file containing a list of $k$-mers and their related counters. 
This textual file can be obtained for a database by KMC-dump program, that is presented in Section~\ref{sec:examples} as a sample application of our KMC API.
Section~\ref{sec:format} describes the database format in detail for those interested in the low-level access to the data.
Section~\ref{sec:results} contains additional experimental results and a description of the parameters of execution of the examined programs.
Section~\ref{sec:params} contains description of how the automatic setting of parameters of KMC works.

As this document is in part a technical documentation of KMC, some parts of it (e.g., API, command-line parameters) are highly similar to the supplement of our previous paper: S.\ Deorowicz, A.\ Debudaj-Grabysz, Sz.\ Grabowski: Disk-based $k$-mer counting on a PC, {\em BMC Bioinformatics}, {\bf} 14, 160 (2013).
The mentioned paper described the previous version of the current tool, i.e., KMC~1.

Below we describe in detail the parameters and options of the KMC command-line tool, in version 2.0.

The general syntax is: \\
kmc [options] $<$input\_file\_name$>$ $<$output\_file\_name$>$ $<$working\_directory$>$ \\
or: \\
kmc [options] $<$@input\_file\_names$>$ $<$output\_file\_name$>$ $<$working\_directory$>$ \\

where the parameters are:

\begin{itemize}
\item input\_file\_name --- a single file in FASTQ format (gzipped or not),
\item @input\_file\_names --- a file name with list of input files in FASTQ format (gzipped or not),
\item output\_file\_name --- the output database file; if such a file exists, it will be overwritten.
\end{itemize}

The configuration options comprise:

\begin{itemize}
\item \textsf{-v} --- verbose mode (shows all parameter settings); default: false,
\item \textsf{-k$<$len$>$} --- $k$-mer length, $k$ from 1 to MAX\_K; default: 25,
\item \textsf{-m$<$size$>$} --- max amount of RAM in GB (from 4 to 1024); default: 12,
\item \textsf{-p$<$par$>$} --- set signature length (from 5 to 7); default: 7,
\item \textsf{-f[a/q/m]} --- input in FASTA format (-fa), FASTQ format (-fq) or multi FASTA (-fm); default: FASTQ,
\item \textsf{-q[value]} --- use Quake's compatible counting with [value] representing lowest quality; default: 33,
\item \textsf{-ci$<$value$>$} --- exclude $k$-mers occurring less than $<$value$>$ times; default: 2,
\item \textsf{-cs$<$value$>$} --- maximal value of a counter; default: 255,
\item \textsf{-cx$<$value$>$} --- exclude $k$-mers occurring more of than $<$value$>$ times; default: 1e9,
\item \textsf{-b} --- turn off transformation of $k$-mers into canonical form,
\item \textsf{-r} --- turn on RAM-only mode,
\item \textsf{-sf$<$value$>$} --- number of FASTQ reading threads,
\item \textsf{-sp$<$value$>$} --- number of splitting threads,
\item \textsf{-so$<$value$>$} --- number of sorter threads,
\item \textsf{-sr$<$value$>$} --- number of threads per single sorter,
\item \textsf{-t$<$value$>$} --- total number of threads.
\end{itemize}

The parameters \textsf{-sf$<$value$>$}, \textsf{-sp$<$value$>$}, \textsf{-so$<$value$>$}, 
and \textsf{-sr$<$value$>$} concern the internal work of KMC, i.e., their settings may affect the program's 
processing speed, but won't change its output.
Not setting at least one parameter from this group makes KMC ignore them all.
The parameter \textsf{-t$<$value$>$} sets total number of threads (including fastq readers, splitters, sorters and threads per single sorter, but NOT including disk writer, bin reader and main KMC thread). If \textsf{-t$<$value$>$}, \textsf{-sf$<$value$>$}, \textsf{-sp$<$value$>$}, \textsf{-so$<$value$>$}, 
and \textsf{-sr$<$value$>$} are specified the \textsf{-t$<$value$>$} is ignored.
Setting the parameter \textsf{-r} causes all computations are performed using RAM memory, without using disk space (memory usage may exceed limit). \\
\\
Here are some usage examples.

kmc -k27 -m24 NA19238.fastq NA.res $\backslash$data$\backslash$kmc\_tmp\_dir$\backslash$

kmc -k27 -q -m24 @files.lst NA.res $\backslash$data$\backslash$kmc\_tmp\_dir$\backslash$

\clearpage
\section{API}\label{sec:api}

In this section we describe two classes, CKmerAPI and CKMCFile.
They can be used to obtain access to the databases produced by KMC program.

\subsection{\textsf{CKmerAPI} class}

This class represents a $k$-mer.
Its key methods are:

\begin{itemize}
\item \textsf{CKmerAPI(uint32 length = 0)} --- constructor, that creates the array kmer\_data of appropriate size,
\item \textsf{CKmerAPI(const CKmerAPI \&kmer)} --- copy constructor,
\item \textsf{char get\_symbol(unsigned int pos)} --- returns $k$-mer's symbol at a given position (0-based),
\item \textsf{std::string to\_string()} --- converts $k$-mer to string, using the alphabet ACGT,
\item \textsf{void to\_string(char *str)} --- converts $k$-mer to string, using the alphabet ACGT; 
   the function assumes that enough memory was allocated,
\item \textsf{void to\_string(str::string \&str)} --- converts $k$-mer to string, using the alphabet ACGT, 
\item \textsf{bool from\_string(std::string \&str)} --- converts string (from alphabet ACGT) to $k$-mer,
\item \textsf{~CKmerAPI()} --- destructor, releases the content of \textsf{kmer\_data} array,
\item overloaded operators: =, ==, $<$.
\end{itemize}

\subsection{\textsf{CKMCFile} class}

This class handles a $k$-mer database.
Its key methods are:

\begin{itemize}
\item \textsf{CKMCFile()} --- constructor,
\item \textsf{bool OpenForRA(std::string file\_name)} --- opens two files: file\_name with added 
   extension ``.kmc\_pre'' and ``.kmc\_suf'', reads their whole content to enable random access (in memory), 
   and then closes them,
\item \textsf{bool OpenForListing(std::string file\_name)} --- opens the file file\_name with added 
   extension ``.kmc\_pre'' and allows to read the $k$-mers one by one (whole database is not loaded into memory),
\item \textsf{bool ReadNextKmer(CKmerAPI \&kmer, float \&count)} --- reads next $k$-mer to kmer and updates its count; 
   the return value is bool; true as long as not eof-of-file (available only when database is opened in listing mode),
\item \textsf{bool Close()} --- if the file was opened for random access, the allocated memory for its 
   content is released; if the file was opened for listing, the allocated memory for its content is released 
   and the ``.kmer'' file is closed,
\item \textsf{bool SetMinCount(uint32 x)} --- set the minimum counter value for $k$-mers; if a $k$-mer has count below x, 
   it is treated as non-existent,
\item \textsf{uint32 GetMinCount(void)} --- returns the value (uint32) set with SetMinCount,
\item \textsf{bool SetMaxCount(uint32 x)} --- set the maximum counter value for $k$-mers; if a $k$-mer has count above x,
   it is treated as non-existent,
\item \textsf{uint32 GetMaxCount(void)} --- returns the value (uint32) set with SetMaxCount,
\item \textsf{uint64 KmerCount(void)} --- returns the number of $k$-mers in the database (available only for databases opened in random access mode),
\item \textsf{uint32 KmerLength(void)} --- returns the $k$-mer length in the database (available only for databases opened in random access mode),
\item \textsf{bool RestartListing(void)} --- sets the cursor for listing $k$-mers from the beginning of the file (available only for databases opened in listing mode).
   The method \textsf{OpenForListing(std::string file\_name)} invokes it automatically, but it can be 
   also called by a user,
\item \textsf{bool Eof(void)} --- returns true if all $k$-mers have been listed,
\item \textsf{bool CheckKmer(CKmerAPI \&kmer, float \&count)} --- returns true if kmer exists in the database 
   and set its count if the answer is positive (available only for databases opened in random access mode),
\item \textsf{bool IsKmer(CKmerAPI \&kmer)} --- returns true if kmer exists (available only for databases opened in random access mode),
\item \textsf{void ResetMinMaxCounts(void)} --- sets min\_count and max\_count to the values read from the database,
\item \textsf{bool Info(uint32 \&\_kmer\_length, uint32 \&\_mode, uint32 \&\_counter\_size, 
   uint32 \&\_lut\_prefix\_length, uint32 \&\_signature\_len, uint32 \&\_min\_count, uint32 \&\_max\_count, uint64 \&\_total\_kmers)} ---
   gets current parameters from the $k$-mer database,
\item \textsf{~CKMCFile()} --- destructor.
\end{itemize}

\clearpage
\section{Example of API usage}\label{sec:examples}

The \textsf{kmc\_dump} application (Figs.~\ref{fig:dump:1} and~\ref{fig:dump:2}) shows how to list and print $k$-mers 
with at least $\mathit{min\_count}$ and at most $\mathit{max\_count}$ occurrences in the database.
Fig.~\ref{fig:dump:1} presents parsing the command-line parameters, including \textsf{-ci$<$value$>$} and
 \textsf{-cx$<$value$>$}.
Input and output file names are also expected.
The code in Fig.~\ref{fig:dump:2} is for actual database handling.
This database is represented by a \textsf{CKMCFile} object, which opens an input file for $k$-mer listing
(the method \textsf{bool OpenForListing(std::string file\_name)} is invoked).
The parameter of the method \textsf{SetMinCount} (\textsf{SetMaxCount}) must be not smaller (not greater) 
than the corresponding parameter \textsf{-ci} (\textsf{-cx}) with which KMC was invoked 
(otherwise, nothing will be listed).
The listed $k$-mers are in the form like:\\
\textsf{AAACACCGT\textbackslash t$<$value$>$}\\
where the first part is the $k$-mer in natural representation,
which is followed by a tab character, and its associated value (integer or float).
(Such format is compatible with Quake, a widely used tool for sequencing error correction.)
Note that, if needed, one can easily modify the output format, changing the lines 39 and 41 
in Fig.~\ref{fig:dump:2}. 

For performance reasons, the KMC package contains two variants of the dump program. 
The first one, presented below, is the \textsf{kmc\_dump\_sample} program.
The second variant, \textsf{kmc\_dump}, is essentially the same, the only difference is the way the counters are printed.
Instead of the \textsf{fprintf} function we used much faster way of converting numbers into the textual form.
Thus, in real applications the \textsf{kmc\_dump} variant should be used.

\begin{figure}
\linespread{1.0}
\begin{lstlisting}
#include <iostream>
#include "../kmc_api/kmc_file.h"

void print_info(void);

int _tmain(int argc, char* argv[])
{
	CKMCFile kmer_database;
	int i;
	uint32 min_count_to_set = 0;
	uint32 max_count_to_set = 0;
	std::string input_file_name;
	std::string output_file_name;

	FILE * out_file;
	//------------------------------------------------------------
	// Parse input parameters
	//------------------------------------------------------------
	if(argc < 3)
	{
		print_info();
		return EXIT_FAILURE;
	}

	for(i = 1; i < argc; ++i)
	{
		if(argv[i][0] == '-')
		{	
			if(strncmp(argv[i], "-ci", 3) == 0)
				min_count_to_set = atoi(&argv[i][3]);
			else if(strncmp(argv[i], "-cx", 3) == 0)
					max_count_to_set = atoi(&argv[i][3]);
		}
		else
			break;
	}

	if(argc - i < 2)
	{ 
		print_info();
		return EXIT_FAILURE;
	}

	input_file_name = std::string(argv[i++]);
	output_file_name = std::string(argv[i]);

	if((out_file = fopen (output_file_name.c_str(),"wb")) == NULL)
	{
		print_info();
		return EXIT_FAILURE;
	}

	setvbuf(out_file, NULL ,_IOFBF, 1 << 24);

	...
\end{lstlisting}
\caption{First part of \textsf{kmc\_dump\_sample} application}
\label{fig:dump:1}
\end{figure}
	
\begin{figure}
\linespread{1.0}
\begin{lstlisting}
	//------------------------------------------------------------------------------
	// Open kmer database for listing and print kmers within min_count and max_count
	//------------------------------------------------------------------------------

	if(!kmer_database.OpenForListing(input_file_name))
	{
		print_info();
		return EXIT_FAILURE ;
	}
	else
	{
		uint32 _kmer_length;
		uint32 _mode;
		uint32 _counter_size;
		uint32 _lut_prefix_length;
		uint32 _signature_len;
		uint32 _min_count;
		uint32 _max_count;
		uint64 _total_kmers;

		kmer_database.Info(_kmer_length, _mode, _counter_size, _lut_prefix_length, _signature_len, 
			_min_count, _max_count, _total_kmers);

		float counter;
		std::string str;
		
		CKmerAPI kmer_object(_kmer_length);
		
		if(min_count_to_set)
			if(!(kmer_database.SetMinCount(min_count_to_set)))
				return EXIT_FAILURE;
		if(max_count_to_set)
			if(!(kmer_database.SetMaxCount(max_count_to_set)))
				return EXIT_FAILURE;	

		while(kmer_database.ReadNextKmer(kmer_object, counter))
		{
			kmer_object.to_string(str);

			if(_mode)		
				fprintf(out_file, "%s\t%f\n", str.c_str(), counter);
			else
				fprintf(out_file, "%s\t%d\n", str.c_str(), (int)counter);
		}
	
		fclose(out_file);
	}

	return EXIT_SUCCESS; 
}

// -------------------------------------------------------------------------
// Print execution options 
// -------------------------------------------------------------------------
void print_info(void)
{
	std::cout << "KMC dump ver. " << KMC_VER << " (" << KMC_DATE << ")\n";
	std::cout << "\nUsage:\nkmc_dump [options] <kmc_database> <output_file>\n";
	std::cout << "Parameters:\n";
	std::cout << "<kmc_database> - kmer_counter's output\n";
	std::cout << "Options:\n";
	std::cout << "-ci<value> - print k-mers occurring less than <value> times\n";
	std::cout << "-cx<value> - print k-mers occurring more of than <value> times\n";
};
\end{lstlisting}
\caption{Second part of \textsf{kmc\_dump\_sample} application}
\label{fig:dump:2}
\end{figure}

\clearpage
\section{Database format}\label{sec:format}

The KMC application creates output files with two extensions: 
\begin{itemize}
\item \textsf{.kmc\_pre} --- with information on $k$-mer prefixes (plus some other data),
\item \textsf{.kmc\_suf} --- with information on $k$-mer suffixes and the related counters.
\end{itemize}

All integers in the KMC output files are stored in LSB (least significant byte first) byte order.

\subsection{The \textsf{.kmc\_pre} file structure}

The \textsf{.kmc\_pre} file contains, in order, the following data:
\begin{itemize}
\item \textsf{[marker]},
\item \textsf{[prefixes]},
\item \textsf{[map]},
\item \textsf{[header]},
\item \textsf{[header position]},
\item \textsf{[marker]} (another copy, to signal the file is not truncated).
\end{itemize}

\subsubsection*{[marker]}
4 bytes with the letters: KMCP.

\subsubsection*{[header position]}
The integer consisting of the last 4 bytes in the file (before end KMCP marker).
It contains the relative position of the beginning of the field [header].
After opening the file, one should do the following:
\begin{enumerate}
\item Read the first 4 bytes and check if they contain the letters KMCP.
\item Read the last 4 bytes and check if they contain the letters KMCP.
\item Jump to position 8 bytes back from end of file and read the header position $x$.
\item Jump to position $x + 8$ bytes back from end of file and read the header.
\item Read [data].
\end{enumerate}

\subsubsection*{[header]}
The header contains fields describing the file .kmc\_pre:
\begin{itemize}
\item \textsf{uint32 kmer\_length} --- $k$-mer length,
\item \textsf{uint32 mode} --- mode: 0 (occurrence counters) or 1 (quality-aware counters),
\item \textsf{uint32 counter\_size} --- counter field size: for mode 0 it is 1, 2, 3, or 4; 
  for mode 1 it is always 4,
\item \textsf{uint32 lut\_prefix\_length} --- the length (in symbols) of the prefix cut off from $k$-mers;
  it is invariant of the scheme that 4 divides $(\mathit{kmer\_length} - \mathit{lut\_prefix\_length})$,
\item \textsf{uint32 signature\_length} --- the length (in symbols) of the signature,
\item \textsf{uint32 min\_count} --- minimum number of $k$-mer occurrences to write in the database 
  (if the counter is smaller, the $k$-mer data are not written),
\item \textsf{uint32 max\_count} --- maximum number of $k$-mer occurrences to write in the database 
  (if the counter is greater, the $k$-mer data are not written),
\item \textsf{uint64 total\_kmers} --- total number of $k$-mers in the database,
\item \textsf{uint32 tmp[7]} --- not used in the current version,
\item \textsf{uint32 KMC\_VER} --- version of KMC software (for KMC 2 this value is equal to 0x200).
\end{itemize}

\subsection*{[map]}
There is an array of \textsf{uint32} elements, of size $4^\mathit{signature\_length} + 1$. This array is used to identify position of proper prefixes' array stored in \textsf{[prefixes]} region. For example, if the queried $k$-mer is \textsf{ATACGACAAATG} and $\mathit{signature\_length} = 5$, its signature is \textsf{ACGAC} (as it is the smallest $5$-mer which satisfies conditions of being a~signature). DNA symbols are encoded as follows: $\mathsf{A} \rightarrow 0$, $\mathsf{C} \rightarrow 1$, $\mathsf{G} \rightarrow 2$, $\mathsf{T} \rightarrow 3$, so \textsf{ACGAC} is equal to 97 (since $0 \cdot 2^8 + 1 \cdot 2^6 + 2 \cdot 2^4 + 0 \cdot 2^2 + 1 \cdot 2^0 = 97 $). In this case we look into ``map'' at position 97 to get the id of related prefixes' array.  

\subsection*{[prefixes]}
This region contains a number of prefixes' arrays (typically hundreds of them) of \textsf{uint64} elements. Each array is of size $4^\mathit{lut\_prefix\_length}$. The last  prefixes' array is followed by an additional \textsf{uint64} element being a~guard to make the reading process simpler. The total number of prefixes' arrays can be easily calculated (as start and end position are given, size of one array is also known). The element at position $x$ in prefixes' array for given signature~$s$ points to a record in .kmc\_suf file. This record contains the first suffix of $k$-mer with prefix $x$ and signature $s$ (the position of the last record can be obtained by decreasing the value at $x+1$ in prefixes' array by~1). 

Using the example from the previous section, the start position of prefixes' array for $k$-mer \textsf{ATACGACAAATG} should be calculated as:
$4 + 97 \cdot 4^\mathit{lut\_prefix\_length} \cdot 8$ (marker $+$ equivalent of \textsf{ACGAC} signature $\cdot$ no.\ of elements in each array $\cdot$ size of element in prefix array). The next step is to cut off the prefix of length equal to $\mathit{lut\_prefix\_length}$ from the queried $k$-mer. Let us assume $\mathit{lut\_prefix\_length} = 4$, and then the prefix is \textsf{ATAC} whose equivalent is 49.
The element at position $49$ in the related prefixes' array (pointed by signature $97$) is the position of the first record in \textsf{.kmc\_suf} file which contains a $k$-mer with prefix \textsf{ATAC} and with signature \textsf{ACGAC}. Let us suppose this position is $1523$, then we look at position $50$ in prefixes' array (say, it contains 1685). 
This means that \textsf{.kmc\_suf} file stores the suffixes of $k$-mers with prefix \textsf{ATAC} and signature \textsf{ACGAC} in the records from $1523$ to $1685-1$. Having got this range, we can now apply binary search for the suffix \textsf{GACAAATG}.

\subsection{The\textsf{.kmc\_suf} file structure}

The \textsf{.kmc\_suf} file contains, in order, the following data:
\begin{itemize}
\item \textsf{[marker]},
\item \textsf{[data]},
\item \textsf{[marker]} (another copy, to signal the file is not truncated).
\end{itemize}

The $k$-mers are stored with their leftmost symbol first, packed into bytes.
For example, \textsf{CCACAAAT} is represented as \textsf{0x51} (for \textsf{CCAC}), \textsf{0x03} (for \textsf{AAAT}).
Integers are stored according to the LSB (little endian) 
byte order,
floats are stored in the same way as they are stored in the memory.

\subsubsection*{[marker]}
4 bytes with the letters: 
KMCS.

\subsubsection*{[data]}
An array \textsf{record\_t records[total\_kmers]}.


\textsf{total\_kmers} is taken from the \textsf{.kmc\_pre} file.

\textsf{record\_t} is a type representing a $k$-mer.
Its first field is the $k$-mer suffix string, stored on $(\mathit{kmer\_length} - \mathit{lut\_prefix\_length}) / 4$ bytes.
The next field is $\mathit{counter\_size}$, with the number of bytes used by the counter, 
which is either a $1 \ldots 4$-byte integer, or a $4$-byte float.

\clearpage
\section{Experimental results}
\label{sec:results}

\subsection{Test platforms}
K-mer Counter (KMC), was implemented in C++11, using gcc compiler (version 4.8.3) for the linux build
and Microsoft Visual Studio 2013 for the Windows build. 

The configuration of the test machine was:
\begin{itemize}
\item CPU: Intel i7 4930 (6-cores clocked at 3.4\,GHz),
\item RAM: 64\,GB RAM (clocked at 1833\,MHz),
\item HDD: 2 drives Seagate Constellation ES.3 3\,TB each in RAID 0; buffered transfers reported by \textsf{hdparm -t}: 355\,MB/s. 
\item SSD: Samsung 840 Evo 1\,TB; buffered transfers reported by \textsf{hdparm -t}: 510\,MB/s.
\end{itemize}

\subsection{Datasets}
\subsubsection{\emph{F.\ vesca}}
The files were downloaded from the following URLs:\\
\url{ftp://ftp.ddbj.nig.ac.jp/ddbj_database/dra/fastq/SRA020/SRA020125/SRX030576/SRR072006.fastq.bz2}\\
\url{ftp://ftp.ddbj.nig.ac.jp/ddbj_database/dra/fastq/SRA020/SRA020125/SRX030576/SRR072007.fastq.bz2}\\
\url{ftp://ftp.ddbj.nig.ac.jp/ddbj_database/dra/fastq/SRA020/SRA020125/SRX030577/SRR072008.fastq.bz2}\\
\url{ftp://ftp.ddbj.nig.ac.jp/ddbj_database/dra/fastq/SRA020/SRA020125/SRX030577/SRR072009.fastq.bz2}\\
\url{ftp://ftp.ddbj.nig.ac.jp/ddbj_database/dra/fastq/SRA020/SRA020125/SRX030578/SRR072013.fastq.bz2}\\
\url{ftp://ftp.ddbj.nig.ac.jp/ddbj_database/dra/fastq/SRA020/SRA020125/SRX030578/SRR072014.fastq.bz2}\\
\url{ftp://ftp.ddbj.nig.ac.jp/ddbj_database/dra/fastq/SRA020/SRA020125/SRX030578/SRR072029.fastq.bz2}\\
\url{ftp://ftp.ddbj.nig.ac.jp/ddbj_database/dra/fastq/SRA020/SRA020125/SRX030575/SRR072005.fastq.bz2}\\
\url{ftp://ftp.ddbj.nig.ac.jp/ddbj_database/dra/fastq/SRA020/SRA020125/SRX030575/SRR072010.fastq.bz2}\\
\url{ftp://ftp.ddbj.nig.ac.jp/ddbj_database/dra/fastq/SRA020/SRA020125/SRX030575/SRR072011.fastq.bz2}\\
\url{ftp://ftp.ddbj.nig.ac.jp/ddbj_database/dra/fastq/SRA020/SRA020125/SRX030575/SRR072012.fastq.bz2}

Then they were decompressed to a single \url{fv.fastq} file.

\subsubsection{\emph{G.\ gallus}}
The files were downloaded from the following URLs:\\
\url{ftp://ftp.ddbj.nig.ac.jp/ddbj_database/dra/fastq/SRA030/SRA030308/SRX043656/SRR105788_1.fastq.bz2}\\
\url{ftp://ftp.ddbj.nig.ac.jp/ddbj_database/dra/fastq/SRA030/SRA030308/SRX043656/SRR105788_2.fastq.bz2}\\
\url{ftp://ftp.ddbj.nig.ac.jp/ddbj_database/dra/fastq/SRA030/SRA030309/SRX043656/SRR105789_1.fastq.bz2}\\
\url{ftp://ftp.ddbj.nig.ac.jp/ddbj_database/dra/fastq/SRA030/SRA030309/SRX043656/SRR105789_2.fastq.bz2}\\
\url{ftp://ftp.ddbj.nig.ac.jp/ddbj_database/dra/fastq/SRA030/SRA030312/SRX043656/SRR105792_1.fastq.bz2}\\
\url{ftp://ftp.ddbj.nig.ac.jp/ddbj_database/dra/fastq/SRA030/SRA030312/SRX043656/SRR105792_2.fastq.bz2}\\
\url{ftp://ftp.ddbj.nig.ac.jp/ddbj_database/dra/fastq/SRA030/SRA030314/SRX043656/SRR105794.fastq.bz2}\\
\url{ftp://ftp.ddbj.nig.ac.jp/ddbj_database/dra/fastq/SRA030/SRA030314/SRX043656/SRR105794_1.fastq.bz2}\\
\url{ftp://ftp.ddbj.nig.ac.jp/ddbj_database/dra/fastq/SRA030/SRA030314/SRX043656/SRR105794_2.fastq.bz2}\\
\url{ftp://ftp.ddbj.nig.ac.jp/ddbj_database/dra/fastq/SRA036/SRA036382/SRX043656/SRR197985.fastq.bz2}\\
\url{ftp://ftp.ddbj.nig.ac.jp/ddbj_database/dra/fastq/SRA036/SRA036382/SRX043656/SRR197985_1.fastq.bz2}\\
\url{ftp://ftp.ddbj.nig.ac.jp/ddbj_database/dra/fastq/SRA036/SRA036382/SRX043656/SRR197985_2.fastq.bz2}\\
\url{ftp://ftp.ddbj.nig.ac.jp/ddbj_database/dra/fastq/SRA036/SRA036383/SRX043656/SRR197986.fastq.bz2}\\
\url{ftp://ftp.ddbj.nig.ac.jp/ddbj_database/dra/fastq/SRA036/SRA036383/SRX043656/SRR197986_1.fastq.bz2}\\
\url{ftp://ftp.ddbj.nig.ac.jp/ddbj_database/dra/fastq/SRA036/SRA036383/SRX043656/SRR197986_2.fastq.bz2}

Then they were decompressed to a single \url{gg.fastq} file.
The files were also re-compressed to gzip format for the experiments with $k$-mer counting of gzipped files.

\subsubsection{\emph{M. balbisiana}}
The files were downloaded from the following URLs:\\
\url{ftp://ftp.ddbj.nig.ac.jp/ddbj_database/dra/fastq/SRA098/SRA098922/SRX339427/SRR956987.fastq.bz2}\\
\url{ftp://ftp.ddbj.nig.ac.jp/ddbj_database/dra/fastq/SRA098/SRA098922/SRX339427/SRR957627.fastq.bz2}

Then they were decompressed to a single \url{mb.fastq} file.

\subsubsection{\emph{H.\ sapiens} 1}
The files were downloaded from the following URL:\\
\url{ftp://ftp.1000genomes.ebi.ac.uk/vol1/ftp/data/HG02057/sequence_read/}

Then they were decompressed to a single \url{hs1.fastq} file.

\subsubsection{\emph{H.\ sapiens} 2}
The FASTQ files (48 files) were downloaded from the following URL:\\
\url{http://www.ebi.ac.uk/ena/data/view/ERA015743}

Then they were decompressed to a single \url{hs2.fastq} file.
The file \url{hs2_files} contains list of gzipped files of this individual.

\subsection{Parameters of programs}
\subsubsection*{Jellyfish}
Jellyfish (ver. 2.1.3) requires to give as a parameter the expected number of counted $k$-mers.
In all experiments we set this value to be about 10\% larger than the number of $k$-mers reported by KMC.

Command lines:\\
\verb"./jellyfish count -m 28 -C -s 300M -t 12 -L 2 -o jelly2 fv.fastq"\\
\verb"./jellyfish count -m 55 -C -s 400M -t 12 -L 2 -o jelly2 fv.fastq"\\
\verb"./jellyfish count -m 28 -C -s 1200M -t 12 -L 2 -o jelly2 gg.fastq"\\
\verb"./jellyfish count -m 55 -C -s 1200M -t 12 -L 2 -o jelly2 gg.fastq"\\
\verb"./jellyfish count -m 28 -C -s 1G -t 12 -L 2 -o jelly2 mb.fastq"\\
\verb"./jellyfish count -m 55 -C -s 1200M -t 12 -L 2 -o jelly2 mb.fastq"\\
\verb"./jellyfish count -m 28 -C -s 3G -t 12 -L 2 -o jelly2 hs1.fastq"\\
\verb"./jellyfish count -m 55 -C -s 3G -t 12 -L 2 -o jelly2 hs1.fastq"\\
\verb"./jellyfish count -m 28 -C -s 3G -t 12 -L 2 -o jelly2 hs2.fastq"\\
\verb"./jellyfish count -m 55 -C -s 3G -t 12 -L 2 -o jelly2 hs2.fastq"\\

\subsubsection*{KAnalyze}
KAnalyze (ver.\ 0.9.5) does not allow to count $k$-mers for $k>32$, so only a single value of $k$ was used in the tests.
Since KAnalyze documentation does not say how to divide the threads among ``k-mer step'' and ``split step'' we allocated 6 threads for both steps (\verb"-l" and \verb"-d" parameters).

Command lines:\\
\verb"java -jar ./kanalyze.jar count -d 6 -f fastq -k 28 -l 6 fv.fastq"\\
\verb"java -jar ./kanalyze.jar count -d 6 -f fastq -k 28 -l 6 gg.fastq"\\
\verb"java -jar ./kanalyze.jar count -d 6 -f fastq -k 28 -l 6 mb.fastq"\\
\verb"java -jar ./kanalyze.jar count -d 6 -f fastq -k 28 -l 6 hs1.fastq"\\
\verb"java -jar ./kanalyze.jar count -d 6 -f fastq -k 28 -l 6 hs2.fastq"\\

\subsubsection*{DSK}
DSK (ver.\ 1.6066) was executed with default parameters that means 6\,GB limit of RAM.
The $k$-mers occurring less than 2 times were excluded.

Command lines:\\
\verb"./dsk32 fv.fastq 28 -t 2 -m 6144 -o o_dsk"\\
\verb"./dsk64 fv.fastq 55 -t 2 -m 6144 -o o_dsk"\\
\verb"./dsk32 gg.fastq 28 -t 2 -m 6144 -o o_dsk"\\
\verb"./dsk64 gg.fastq 55 -t 2 -m 6144 -o o_dsk"\\
\verb"./dsk32 mb.fastq 28 -t 2 -m 6144 -o o_dsk"\\
\verb"./dsk64 mb.fastq 55 -t 2 -m 6144 -o o_dsk"\\
\verb"./dsk32 hs1.fastq 28 -t 2 -m 6144 -o o_dsk"\\
\verb"./dsk64 hs1.fastq 55 -t 2 -m 6144 -o o_dsk"\\
\verb"./dsk32 hs2.fastq 28 -t 2 -m 6144 -o o_dsk"\\
\verb"./dsk64 hs2.fastq 55 -t 2 -m 6144 -o o_dsk"\\

\subsubsection*{Turtle}
The program scTurtle (ver.\ 0.3) was used to calculate the $k$-mers and their counts.
The documentation says that the number of threads should be a prime, so for our 12-virtual cores system we used 11 threads.
The expected number of $k$-mers was set to be about 10\% larger than the exact value (calculated by KMC). 

Command lines:\\
\verb"./scTurtle32 -f fv.fastq -o turtle -k 28 -t 11 -n 400000000"\\
\verb"./scTurtle64 -f fv.fastq -o turtle -k 55 -t 11 -n 400000000"\\
\verb"./scTurtle32 -f gg.fastq -o turtle -k 28 -t 11 -n 1150000000"\\
\verb"./scTurtle64 -f gg.fastq -o turtle -k 55 -t 11 -n 1150000000"\\
\verb"./scTurtle32 -f mb.fastq -o turtle -k 28 -t 11 -n 1100000000"\\
\verb"./scTurtle64 -f mb.fastq -o turtle -k 28 -t 11 -n 1100000000"\\
\verb"./scTurtle32 -f hs1.fastq -o turtle -k 28 -t 11 -n 3000000000"\\
\verb"./scTurtle64 -f hs1.fastq -o turtle -k 28 -t 11 -n 3000000000"\\
\verb"./scTurtle32 -f hs2.fastq -o turtle -k 28 -t 11 -n 3000000000"\\
\verb"./scTurtle64 -f hs2.fastq -o turtle -k 28 -t 11 -n 3000000000"\\

\subsubsection*{MSPKmerCounter}
MSPKmerCounter (ver.\ 0.10.0) was used with minimizer length (10) and number of bins (1000) suggested in the original paper.

Command lines:\\
\verb"java -jar ./Partition.jar -in fv.fastq -k 28 -L 353 -NB 1000 -p 10 -t 12"\\
\verb"java -jar ./Count32.jar -t 12 -k 28 -NB 1000"\medskip\\
\verb"java -jar ./Partition.jar -in fv.fastq -k 55 -L 353 -NB 1000 -p 10 -t 12"\\
\verb"java -jar ./Count64.jar -t 12 -k 55 -NB 1000"\medskip\\
\verb"java -jar ./Partition.jar -in gg.fastq -k 28 -L 100 -NB 1000 -p 10 -t 12"\\
\verb"java -jar ./Count32.jar -t 12 -k 28 -NB 1000"\medskip\\
\verb"java -jar ./Partition.jar -in gg.fastq -k 55 -L 100 -NB 1000 -p 10 -t 12"\\
\verb"java -jar ./Count64.jar -t 12 -k 55 -NB 1000"\medskip\\
\verb"java -jar ./Partition.jar -in mb.fastq -k 28 -L 101 -NB 1000 -p 10 -t 12"\\
\verb"java -jar ./Count32.jar -t 12 -k 28 -NB 1000"\medskip\\
\verb"java -jar ./Partition.jar -in mb.fastq -k 55 -L 101 -NB 1000 -p 10 -t 12"\\
\verb"java -jar ./Count64.jar -t 12 -k 55 -NB 1000"\medskip\\
\verb"java -jar ./Partition.jar -in hs1.fastq -k 28 -L 100 -NB 1000 -p 10 -t 12"\\
\verb"java -jar ./Count32.jar -t 12 -k 28 -NB 1000"\medskip\\
\verb"java -jar ./Partition.jar -in hs1.fastq -k 55 -L 100 -NB 1000 -p 10 -t 12"\\
\verb"java -jar ./Count64.jar -t 12 -k 55 -NB 1000"\medskip\\
\verb"java -jar ./Partition.jar -in hs2.fastq -k 28 -L 101 -NB 1000 -p 10 -t 12"\\
\verb"java -jar ./Count32.jar -t 12 -k 28 -NB 1000"\medskip\\
\verb"java -jar ./Partition.jar -in hs2.fastq -k 55 -L 101 -NB 1000 -p 10 -t 12"\\
\verb"java -jar ./Count64.jar -t 12 -k 55 -NB 1000"\\

\subsubsection*{KMC 1}
KMC was executed with setting that $k$-mers occurring less than 2 times should not be counted.
The parameter \textsf{-p} (prefix length) was set to 5 for all data sets except the smallest one.

Command lines:\\
\verb"./kmc1 -v -m16 -k28 -p4 fv.fastq res temp"\\
\verb"./kmc1 -v -m16 -k55 -p4 fv.fastq res temp"\\
\verb"./kmc1 -v -m16 -k28 -p5 gg.fastq res temp"\\
\verb"./kmc1 -v -m16 -k55 -p5 gg.fastq res temp"\\
\verb"./kmc1 -v -m16 -k28 -p5 mb.fastq res temp"\\
\verb"./kmc1 -v -m16 -k55 -p5 mb.fastq res temp"\\
\verb"./kmc1 -v -m16 -k28 -p5 hs1.fastq res temp"\\
\verb"./kmc1 -v -m16 -k55 -p5 hs1.fastq res temp"\\
\verb"./kmc1 -v -m16 -k28 -p5 hs2.fastq res temp"\\
\verb"./kmc1 -v -m16 -k55 -p5 hs2.fastq res temp"\\

\subsubsection*{KMC 2}
KMC was executed with setting that $k$-mers occurring less than 2 times should not be counted.
The parameter \textsf{-p} (minimizer length) was set to 7 for all data sets.

Command lines (main tests):\\
\verb"./kmc2 -v -m12 -k28 -p7 fv.fastq res temp"\\
\verb"./kmc2 -v -m12 -k28 -p7 fv.fastq res temp"\\
\verb"./kmc2 -v -m6 -k28 -p7 fv.fastq res temp"\\
\verb"./kmc2 -v -m6 -k28 -p7 fv.fastq res temp"\\
\verb"./kmc2 -v -m12 -k28 -p7 gg.fastq res temp"\\
\verb"./kmc2 -v -m12 -k28 -p7 gg.fastq res temp"\\
\verb"./kmc2 -v -m6 -k28 -p7 gg.fastq res temp"\\
\verb"./kmc2 -v -m6 -k28 -p7 gg.fastq res temp"\\
\verb"./kmc2 -v -m12 -k28 -p7 mb.fastq res temp"\\
\verb"./kmc2 -v -m12 -k28 -p7 mb.fastq res temp"\\
\verb"./kmc2 -v -m6 -k28 -p7 mb.fastq res temp"\\
\verb"./kmc2 -v -m6 -k28 -p7 mb.fastq res temp"\\
\verb"./kmc2 -v -m12 -k28 -p7 hs1.fastq res temp"\\
\verb"./kmc2 -v -m12 -k28 -p7 hs1.fastq res temp"\\
\verb"./kmc2 -v -m6 -k28 -p7 hs1.fastq res temp"\\
\verb"./kmc2 -v -m6 -k28 -p7 hs1.fastq res temp"\\
\verb"./kmc2 -v -m12 -k28 -p7 hs2.fastq res temp"\\
\verb"./kmc2 -v -m12 -k28 -p7 hs2.fastq res temp"\\
\verb"./kmc2 -v -m6 -k28 -p7 hs2.fastq res temp"\\
\verb"./kmc2 -v -m6 -k28 -p7 hs2.fastq res temp"\\

Command lines (gzipped files):\\
\verb"./kmc2 -v -m12 -k28 -p7 @hs2_files res temp"\\
\verb"./kmc2 -v -m12 -k55 -p7 @hs2_files res temp"\\
where \verb"hs2_files" contains list of gzipped FASTQ files of \verb"hs2" data set.\\

Command lines (thread tests):\\
\verb"taskset -c 0 ./kmc2 -v -m12 -k28 -p7 -t2 @gg_files res temp"\\
\verb"taskset -c 0 ./kmc2 -v -m12 -k28 -p7 -t2 @gg_files res temp"\\
\verb"./kmc2 -v -m12 -k55 -p7 -t2 @gg_files res temp"\\
\verb"./kmc2 -v -m12 -k28 -p7 -t2 @gg_files res temp"\\
\verb"./kmc2 -v -m12 -k55 -p7 -t2 @gg_files res temp"\\
\verb"./kmc2 -v -m12 -k28 -p7 -t3 @gg_files res temp"\\
\verb"./kmc2 -v -m12 -k55 -p7 -t3 @gg_files res temp"\\
\verb"./kmc2 -v -m12 -k28 -p7 -t4 @gg_files res temp"\\
\verb"./kmc2 -v -m12 -k55 -p7 -t4 @gg_files res temp"\\
\verb"./kmc2 -v -m12 -k28 -p7 -t5 @gg_files res temp"\\
\verb"./kmc2 -v -m12 -k55 -p7 -t5 @gg_files res temp"\\
\verb"./kmc2 -v -m12 -k28 -p7 -t6 @gg_files res temp"\\
\verb"./kmc2 -v -m12 -k55 -p7 -t6 @gg_files res temp"\\
\verb"./kmc2 -v -m12 -k28 -p7 -t8 @gg_files res temp"\\
\verb"./kmc2 -v -m12 -k55 -p7 -t8 @gg_files res temp"\\
\verb"./kmc2 -v -m12 -k28 -p7 -t10 @gg_files res temp"\\
\verb"./kmc2 -v -m12 -k55 -p7 -t10 @gg_files res temp"\\
\verb"./kmc2 -v -m12 -k28 -p7 -t12 @gg_files res temp"\\
\verb"./kmc2 -v -m12 -k55 -p7 -t12 @gg_files res temp"\\
where \verb"gg_files" contains list of gzipped FASTQ files of \verb"gg" data set.

Since KMC 2 does not allow to specify less than 2 threads to measure the speed of KMC for a single thread scenario we allowed to use a single core by using the linux \textsf{taskset} command.

\subsection{Results}
The results for additional data sets (the ones that are not included in the main part of the paper) are given below.

\begin{table}[p]
\caption{$k$-mers counting results for \emph{F.\ vesca}.
MSPKC fails, probably due to the variable length of reads in the dataset.}
\renewcommand{\tabcolsep}{1.0em}
\centering\begin{tabular}{@{\extracolsep{0.45em}}lcccp{0.4em}ccc}
\toprule
& \multicolumn{3}{c}{$k=28$} && \multicolumn{3}{c}{$k=55$}\\\cline{2-4}\cline{6-8}
Algorithm 	& RAM	& Disk& Time		&& RAM& Disk 	& Time	 \\
\midrule
\multicolumn{8}{c}{\bfseries SSD}\\
Jellyfish 2	& \q9	& \q0	&	133		&& 39	&\q0	&	243		\\
KAnalyze		& \q9	& 33	&  345		&&	\mcNS\\
DSK			& \q6	& 12	&	141		&&	\q6&	13	&	298		\\
Turtle		& 17	& \q0	&	133		&&	26	&	\q0&	175		\\
MSPKC			& \mcF						&& \mcF						\\
KMC~1			& 13	& 17	&	\q84		&&	17	& 	41	&  243		\\
KMC~2 (12GB)& \q7 	& \q4	& 	\q45		&& 12	& \q3	&	\q59		\\
KMC~2	(6GB)	& \q6	& \q4	& 	\q33		&&\q6	& 	\q3	&	\q60		\\
\midrule
\multicolumn{8}{c}{\bfseries HDD}\\
Jellyfish 2	& \q9	& \q0	&	133		&& 39	&\q0	&	245		\\
DSK			& \q6	& 12	&	147		&&	\q6&	13	&	308		\\
Turtle		& 17	& \q0	&	135		&&	26	&	\q0&	178		\\
KMC~1			& 11	& 17	&	120		&&	17	&	41	&	245		\\
KMC~2			& \q7	& \q4	&	\q58		&&	12	&	\q3&	\q61		\\
\botrule
\end{tabular}
\end{table}

\begin{table}[p]
\caption{$k$-mers counting results for \emph{H.\ sapiens} 1.}
\renewcommand{\tabcolsep}{1.0em}
\centering\begin{tabular}{@{\extracolsep{0.45em}}lcccp{0.4em}ccc}
\toprule
& \multicolumn{3}{c}{$k=28$} && \multicolumn{3}{c}{$k=55$}\\\cline{2-4}\cline{6-8}
Algorithm 	& RAM	& Disk& Time		&& RAM& Disk 	& Time	 \\
\midrule
\multicolumn{8}{c}{\bfseries SSD}\\
Jellyfish 2	& 62 	&\qq0	&\q2,013		&& \mcOM\\
KAnalyze		& \mcOD		 				&& \mcNS\\
DSK			& \q6	& 	192&\q3,485		&&\q6 &	236&\q4,475		\\
Turtle		& \mcOM						&&	\mcOM\\
MSPKC			& 17	& 286	&	10,032	&& \mcOT\\
KMC~1			& 17	& 251	&\q1,930		&&	17	&426	&\q3,788		\\
KMC~2 (12GB)& 12	& \q64&\q1,010		&& 12	&\q44	&\q1,251		\\
KMC~2	(6GB)	&\q6	& \q64&\q1,013		&&\q8	&\q44	&\q1,397		\\
\midrule
\multicolumn{8}{c}{\bfseries HDD}\\
Jellyfish 2	& 62 	&\qq0	&\q2,209		&& \mcOM\\
DSK			& \q6	&192	&10,667		&&\q6	&236	&13,550		\\
MSPKC			& 17	& 286	&	13,444	&& \mcOT\\
KMC~1			& 17	& 251	&\q3,296		&& 17	& 426	&\q5,136		\\
KMC~2			& 12	&\q64	&\q1,417		&&	12	&\q44	&\q1,651		\\
\botrule
\end{tabular}
\end{table}

\clearpage
\section{Automatic setting of parameters in KMC}
\label{sec:params}

The automatic setting of parameters mechanism tries to allocate the available resources (i.e., CPU cores) in the best possible way.
The optimal number of threads for the parts of the algorithm is, however, hard to obtain, since it depends on many things, like the compression method of input files, the speed of disks, etc.
Thus, our automatic mechanism is obviously suboptimal, nevertheless, experiments show that it performs reasonably well. 
If the results are unsatisfactory, the KMC 2 user can specify these parameters from command line.

The most important factor of the mechanism is the number of available cores 
(possibly overridden if the user specifies it with 
\textsf{-t} parameter). FASTQ readers, splitters, sorters and sorting threads per  single sorter are the most important threads of KMC 2. To set an exact number of those threads, \textbf{all} \textsf{-s?} parameters must be specified (if one is omitted, the rest of them is ignored). 

The automatic setting of parameters for the first stage works as follow. 
If the input files are in plain text format (not compressed), there is one FASTQ reader thread. Otherwise the number of FASTQ reader threads is equal to the number of large input files (``large'' means here the ones whose size is greater than 5\% of the size of the largest file), but not more than half of the number of cores. 
After that the number of ``free'' cores (not assigned yet) is set as the number of splitting threads.
In the second stage the memory requirements for each bin are known and the following steps are performed. Bins are sorted by their size in a non-increasing order. 
The number of sorter threads is calculated as
$\lfloor M_2/B_{10}\rfloor$, where $M_2$ is total amount of memory available for the second stage, $B_{10}$ is the size of the bin for which \textsf{10\%} of bins are bigger. The number of sorting threads per single sorter is equal to the number of cores divided by the number of sorter threads (as the result may not be integer, some sorters have one sorting thread more than others, e.g., if there are 7 sorters and 10 threads to allocate, 3 sorters would run 2 sorting threads and 4 sorters only 1 sorting thread).

\end{document}